\documentclass[aps,preprint,prd,nofootinbib]{revtex4}
\usepackage{graphicx}
\usepackage{psfrag}

\begin{document}

\title{The $B_c$ Decays to $P$-wave Charmonium by
Improved Bethe-Salpeter Approach}
\author{Zhi-hui Wang$^{[1]}$, Guo-Li Wang$^{[1]}$\footnote{gl\_wang@hit.edu.cn},
Chao-Hsi Chang$^{[2,3]}$\footnote{zhangzx@itp.ac.cn}\\}
\address{$^1$Department of Physics, Harbin Institute of
Technology, Harbin, 150001, China.\\
$^2$CCAST (World Laboratory), P.O.Box 8730, Beijing 100190, China.\\
$^3$Institute of Theoretical Physics, Chinese Academy of Sciences,
P.O.Box 2735, Beijing 100190, China.\\}
\baselineskip=20pt

\begin{abstract}

We re-calculate the exclusive semileptonic and nonleptonic decays
of $B_c$ meson to a $P$-wave charmonium in terms of the improved
Bethe-Salpeter (B-S) approach, which is developed recently. Here
the widths for the exclusive semileptonic and nonleptonic decays,
the form factors, and the charged lepton spectrums for the
semileptonic decays are precisely calculated. To test the
concerned approach by comparing with experimental measurements
when the experimental data are available, and to have comparisons
with the other approaches the results obtained by the approach and
those by some approaches else as well as the original B-S
approach, which appeared in literature, are comparatively
presented and discussed.

\vspace{4mm}


\end{abstract}

\maketitle

\section{Introduction}

The meson $B_c$ is the ground state of the double heavy (both of the
components are heavy) quark-antiquark binding system ($\bar{b}c$).
In Stand Model (SM) it is an unique meson which carries two
different heavy flavors explicitly, thus it decays weakly, that is
very different from the ground states of flavor-hidden double heavy
mesons, such as $\eta_c$ and $\eta_b$. Namely $B_c$ decays via weak
interaction (via virtual $W$ emitting or annihilating) only, while
the ground states of flavor-hidden double heavy mesons decay
dominantly by annihilating to gluons (strong interaction) or/and
photons (electronic interaction). The meson $B_c$ has very rich and
experimentally accessible decay channels, so to study the decays of
$B_c$ meson is specially interesting. By comparing the experimental
and theoretical results of the decays of $B_c$ meson, we can also
reach some insight into the binding effects of the heavy
quark-antiquark system, which are of QCD nature, besides the
knowledge of the weak interaction such as the CKM matrix elements
etc.

The meson $B_c$ was first experimentally discovered by the CDF
collaboration at Fermilab through the semileptonic decay $B_c\to
J/\psi+l+\nu$ \cite{Abe}, and soon it is confirmed not only by CDF
itself via another decay channel $B_c\to J/\psi+\pi$ \cite{Abe1},
but also by the other collaboration D0 at Fermilab \cite{d0}. The
latest experimental report for its lifetime and mass in PDG
\cite{pdg} is $M_{B_c}=6.277\pm 0.006$ GeV and
${\tau}_{B_c}=(0.453\pm0.041)\times 10^{-12}$ s. Because the cross
section of $B_c$ production is comparatively small, so to discover
it is quite difficult in experiment. Whereas according to the
estimates \cite{chang,bord,bord1}, that LHC will produce about
$5\times 10^{10} B_c$ events per year, it is expected that more
measurements of decays and production of the meson $B_c$ are
available soon at LHC (LHCb, CMS, ATLAS), and it must push more
studies of the decays of $B_c$ meson forward. So both experimental
and theoretical studies on $B_c$ meson now become more
interesting.

In fact, the decays of $B_c$ meson can be divided into three
categories: i). The anti-bottom quark $\bar{b}$ decays into
$\bar{c}$ (or $\bar{u}$) with $c$-quark being as a spectator;  ii).
The charm quark $c$ decays into $s$ (or $d$) with $\bar{b}$-quark
being as a spectator; iii). The two components, $\bar{b}$ and $c$,
annihilate weakly. According to the decay products we may realize
which one or two even three of the categories play roles in a
concerned decay, thus one can measure the CKM elements such as
$V_{bc}$, $V_{ub}$, $V_{cs}$, $V_{cd}$ through the decays. In the
present paper, we are highlighting the decays of $B_c$ meson to a
$P$-wave charmonium, and one may easily to realize that the decays
being considered here belong to the category i). Since the lepton
spectrum and the weak form factors, which relate to the binding
effects (wave functions) precisely, may be measurable in semi-lepton
decays as long as the experimental sample of decay events is great
enough, so we will share quite a lot of lights on them.

In fact, one may find a lot of theoretical methods to treat the
semi-leptonic and non-leptonic decays of $B_c$ meson, such as the
varieties of relativistic constituent quark models
\cite{vary,chang-D,qwerr,likhoded,Bra,Bww,Bww1,wqd} and QCD sum
rules \cite{Az,Lu} etc in the literature, and moreover one may
realize that among the relativistic constituent quark models, the
method presented in Ref.~\cite{previous} and adopted in
Refs.~\cite{chang-D,qwerr} is based on the instantaneous version
\cite{Salp} of the Bethe-Salpeter (B-S) equation \cite{BS}, and the
`instantaneous treatment' is also extended to the weak-current
matrix elements using the Mandelstam formulation \cite{Mand}, while
the adopted approach in Ref.~\cite{vary} is different from the one
presented in Ref.~\cite{previous} only in the kernel of the B-S
equation and the `instantaneous treatment' etc. Recently in
\cite{improv} an improvement to that of \cite{previous} is proposed,
and the relativistic effects in the binding systems and decays
between the systems  may be considered by the new development more
properly, especially, considering the fact that, of the new
development, the part (factor) for dealing with the binding effects
has been applied to study (test) the spectra of positronium (a QED
binding system) \cite{chenjk} and double heavy flavor binding
systems (QCD binding systems) \cite{spectr} and quite satisfied
results are obtained (see Refs.~\cite{chenjk,spectr}), so to test
the new development \cite{improv} when experimental data are
available in foreseeable future, in this paper we try to apply the
development to the decays of $B_c$ meson to a $P$-wave charmonium
and to compare the obtained results with those obtained by old
method in Ref.~\cite{previous} and obtained by other theoretical
approaches. Since we suspect that the decays of $B_c$ meson to a
$P$-wave charmonium might be more sensitive in testing the effects
caused by the improvement than the decays of $B_c$ meson to an
$S$-wave charmonium, so here we focus our attention on the decays of
$B_c$ meson to a $P$-wave charmonium.

The new development \cite{improv} contains two factors: one is
about relativistic wave functions which describe bound states with
definite quantum numbers, i.e. a relativistic form of wave
functions (see Appendix C) which are solutions of the full
Salpeter equation (see Appendix B). Note that here we solve the
full equations Eqs.~(B9,~B10,~B11), not only the first one
Eq.~(B9) as other authors did. The other factor of the improvement
is about computing the weak-current matrix elements for the decays
with the obtained relativistic wave functions as input. It is more
'complete' than that as done in
Refs.~\cite{chang-D,qwerr,previous}, i.e. the 'complete' formula
in Eq.~(15).

The paper is organized as follows: the formulations of the exclusive
semi-leptonic and non-leptonic decays are outlined in Sec.~II. The
newly developed formulations, mainly for the matrix elements of the
hadron weak decays, are presented in Sec.~III. In Sec.~IV, numerical
calculations for the exclusive semi-leptonic decays and non-leptonic
decays are described, the results and comparisons among the various
approaches are presented. Finally the Sec.~V is attributed to
discussions. In Appendices, the formulations as necessary pieces for
the calculations of the decays are given.

\section{The formulations for exclusive semi-leptonic decays and non-leptonic
decays}

Let us now derive the formulations for the exclusive semi-leptonic
and non-leptonic decays precisely (mainly quoted from
Ref.~\cite{improv}) for numerical calculations later on.

In the following subsections we will focus light on the matrix
elements of weak currents, and show how to present the amplitudes of
the semileptonic or nonleptonic decays via the matrix elements of
weak currents precisely. In fact, one may see that the newly
developed method mainly is about the matrix elements of weak
currents.

\subsection{The semileptonic decays of $B_c$ meson}
The Fig. 1 is a typical Feymann diagram responsible for a
semileptonic decay of $B_c$ meson to a charmonium. The
corresponding amplitude for the decay can be written as:
\begin{figure}[htpb]
\centering
\includegraphics[width = 0.5\textwidth]{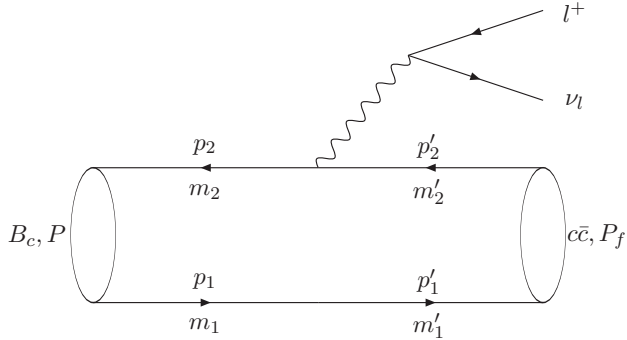}
\caption{The Feynman diagram of a semileptonic decay of $B_c$ meson
to a charmonium.}
\end{figure}
\begin{equation}\label{weak-current}
T=\frac{G_F}{\sqrt2}V_{bc}\bar{u}_{{\nu}_l}(p_\nu){\gamma}_{\mu}
(1-{\gamma}_5){\upsilon}_l(p_l)\langle
\chi_c(h_{c})(P_f)|J^{\mu}|B_c(P)\rangle,
\end{equation}
where $V_{bc}$ is the CKM matrix element, $\langle
\chi_c(h_{c})(P_f)|J^{\mu}|B_c(P)\rangle$ is the hadronic
weak-current matrix element responsible for the decay, and $P$,
$P_f$, $p_\nu$ and $p_l$ are the momenta of initial state $B_c$, the
finial $P$-wave state of $(c\bar{c})$ (i.e. $h_c$, $\chi_{c0}$,
$\chi_{c1}$, $\chi_{c2}$ and their excited states), the neutrino and
the charged lepton respectively.

Generally, the form factors are defined in terms of the matrix
elements of weak current responsible for the decays appearing in
Eq.~(\ref{weak-current}). Namely for the decay of $B_c$ meson to
scalar charmonium $\chi_{c0}$, the form factors $s_+$ and $s_-$
are defined as follows:
\begin{eqnarray}
&&\langle{\chi}_{c0}(P_f)|V^{\mu}|B_c(P)\rangle =0,\nonumber\\&&
\langle{\chi}_{c0}(P_f)|A^{\mu}|B_c(P)\rangle
=s_+(P+P_f)^{\mu}+s_-(P-P_f)^{\mu}\,.
\end{eqnarray}
For the decay of $B_c$ meson to vector charmonium $\chi_{c1}$, the
relevant form factors $f$, $u_1$, $u_2$ and $g$ are defined as
follows:
\begin{eqnarray}
&& \langle {\chi}_{c1}(P_f)|V^{\mu}|B_c(P)\rangle
=f(M+M_f){\varepsilon}^{\mu}+[u_1P^{\mu}+u_2P_f^{\mu}]\frac{{\varepsilon}\cdot
P}{M},\nonumber\\
&& \langle {\chi}_{c1}(P_f)|A^{\mu}|B_c(P)\rangle
=\frac{2g}{M+M_f}i{\epsilon}^{\mu\nu\rho\sigma}
{\varepsilon}_{\nu}P_{\rho}{P_f}_{\sigma}\,.
\end{eqnarray}
For the decay of $B_c$ meson to vector charmonium  $h_c$, the
relevant form factors $V_0$, $V_1$, $V_2$ and $V_3$ are defined as
follows:
\begin{eqnarray}
&& \langle h_c(P_f)|V^{\mu}|B_c(P)\rangle
=V_0(M+M_f){\varepsilon}^{\mu}+[V_1P^{\mu}+V_2P_f^{\mu}]\frac{{\varepsilon}\cdot
P}{M},\nonumber\\
&& \langle h_c(P_f)|A^{\mu}|B_c(P)\rangle
=\frac{2V_3}{M+M_f}i{\epsilon}^{\mu\nu\rho\sigma}
{\varepsilon}_{\nu}P_{\rho}{P_f}_{\sigma}\,.
\end{eqnarray}
For the decay of $B_c$ meson to tenser charmonium $\chi_{c2}$, the
relevant form factors $k$, $c_1$, $c_2$ and $h$ are defined as
follows:
\begin{eqnarray}
&&\langle{\chi}_{c2}(P_f)|A^{\mu}|B_c(P)\rangle
=k(M+M_f){\varepsilon}^{\mu\alpha}\frac{P_{\alpha}}{M}
+{\varepsilon}_{\alpha\beta}\frac{P^{\alpha}P^{\beta}}{M^2}
(c_1P^{\mu}+c_2P_f^{\mu}),\nonumber\\
&& \langle{\chi}_{c2}(P_f)|V^{\mu}|B_c(P)\rangle=
\frac{2h}{M+M_f}i{\varepsilon}_{\alpha\beta}\frac{P^{\alpha}}{M}
{\epsilon}^{\mu\beta\rho\sigma}P_{\rho}{P_f}_{\sigma}\,.
\end{eqnarray}

In the case without considering polarization, we have the squared
decay-amplitude with the polarizations in final states being summed:
\begin{equation}
\Sigma_{s_\nu,s_l,S_{\chi_c(h_c)}}|T|^2=\frac{G_F^2}{2}|V_{bc}|^2l_{\mu\nu}h^{\mu\nu},
\end{equation}
where $l_{\mu\nu}$ is the leptonic tensor:
$$l_{\mu\nu}=\Sigma_{s_\nu,s_l}\bar{{\upsilon}_l}(p_l){\gamma}_{\mu}(1-{\gamma}_5)
{u}_{{\nu}_l}(p_\nu)\bar{u}_{{\nu}_l}(p_\nu){\gamma}_{\nu}
(1-{\gamma}_5){\upsilon}_l(p_l),$$ and the hadronic tensor
relating to the weak-current in Eq.~(\ref{weak-current}) is
\begin{eqnarray}
&h^{\mu\nu} \equiv \Sigma_{S_{\chi_c(h_c)}}\langle
B_c(P)|J^{\mu+}|\chi_c(h_{c})(P_f)\rangle\langle
\chi_c(h_{c})(P_f)|J^{\nu}|B_c(P)\rangle\nonumber\\
& =-{\alpha}g^{\mu\nu}+{\beta}_{++}(P+P_f)^{\mu}(P+P_f)^{\nu}
+{\beta}_{+-}(P+P_f)^{\mu}(P-P_f)^{\nu}\nonumber\\
&
+{\beta}_{-+}(P-P_f)^{\mu}(P+P_f)^{\nu}+{\beta}_{--}(P-P_f)^{\mu}(P-P_f)^{\nu}\nonumber\\
&+i\gamma{\epsilon}^{\mu\nu\rho\sigma}(P+P_f)_{\rho}(P-P_f)_{\sigma},
\end{eqnarray}
where the functions $\alpha$, $\beta_{++}$, $\beta_{+-}$,
$\beta_{-+}$, $\beta_{--}$, $\gamma$ are related to the form factors
and we put the relations in Appendix A precisely.

The total decay width $\Gamma$ can be written as:
\begin{eqnarray}\label{total-g}
\Gamma&=&\frac{1}{2M(2\pi)^9}\int\frac{d^3\vec{P}_f}{2E_f}
\frac{d^3\vec{p}_l}{2E_l}\frac{d^3\vec{p}_{\nu}}{2E_{\nu}}
(2\pi)^4{\delta}^4(P-P_f-p_l-p_{\nu})\Sigma_{s_\nu,s_l,S_{\chi_c(h_c)}}|T|^2,
\end{eqnarray}
where $E_f$, $E_l$ and $E_\nu$ are the energies of the charmonium,
the charged lepton and the neutrino respectively. If we define
$x\equiv E_l/M,\;\; y\equiv (P-P_f)^2/M^2$, the differential width
of the decay can be reduced to:
\begin{eqnarray}\label{differ}
&\displaystyle\frac{d^2\Gamma}{dxdy}=|V_{bc}|^2\frac{G_F^2M^5}{64{\pi}^3}\left\{
\frac{2\alpha}{M^2}(y-\frac{m_l^2}{M^2})\right.\nonumber\\
&\displaystyle+{\beta}_{++}\left[4\left(2x(1-\frac{M_f^2}{M^2}+y)-4x^2-y\right)
+\frac{m_l^2}{M^2}\left(8x+4\frac{M_f^2}{M^2}-3y-\frac{m_l^2}{M^2}\right)\right]\nonumber\\
&\displaystyle+({\beta}_{+-}+{\beta}_{-+})\frac{m_l^2}{M^2}
\left(2-4x+y-2\frac{M_f^2}{M^2}+\frac{m_l^2}{M^2}\right)
+{\beta}_{--}\frac{m_l^2}{M^2}\left(y-\frac{m_l^2}{M^2}\right)\nonumber\\
&\displaystyle\left.-\left[2{\gamma}y\left(1-\frac{M_f^2}{M^2}-4x+y+\frac{M_l^2}{M^2}\right)+
2\gamma\frac{M_l^2}{M^2}\left(1-\frac{M_f^2}{M^2}\right)\right]\right\}\,,
\end{eqnarray}
here $M$ is the mass of the meson $B_c$, $M_f$ is the mass of the
charmonium in final state, and the total width of the decay is just
an integration of the differential width i.e. $\Gamma=\int dx\int
dy\frac{d^2\Gamma}{dxdy}$.

Thus the key problem for calculating the semileptonic decays is
turned to calculating the hadronic weak-current matrix elements.

\subsection{The nonleptonic decays of $B_c$ meson}
\begin{figure}[htpb]
\centering
\includegraphics[width = 0.5\textwidth]{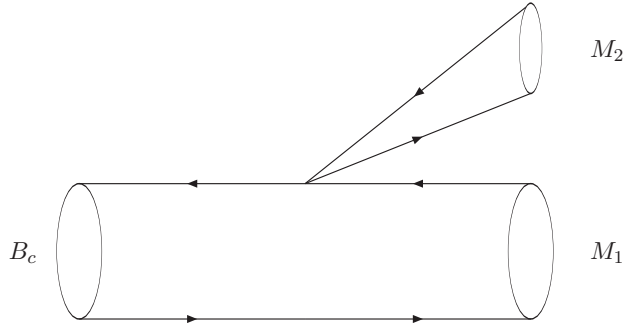}
\caption{The Feynman diagram of a nonleptonic decay of $B_c$ to two
mesons $M_1$ (a charmonium) and $M_2$ (a common meson).}
\end{figure}

In this subsection we mainly consider the nonleptonic two-body
decays to a $P$-wave charmonium, i.e. decays $B_c\to M_1M_2$ where
$M_1$ is a $P$-wave charmonium and $M_2$ is a common meson. Fig.~2
is the Feynman diagram for the decays via the relevant effective
Hamiltonian $H_{eff}$ \cite{mod,ali}:
\begin{eqnarray}
H_{eff}=\frac{G_F}{\sqrt2}\left\{V_{cb}\left[c_1(\mu)O_1^{cb}+c_2(\mu)O_2^{cb}\right]
-V_{tb}V_{tq}^*\left(\sum_{i=3}^{10}C_i(\mu)O_i\right)\right\}+h.c.,
\end{eqnarray}
where $G_F$ is the Fermi constant, $q=d,s$, $V_{ij}$ are the CKM
matrix elements and $c_i(\mu)$ are the scale-dependent Wilson
coefficients. $O_i$ are the operators constructed by four quark
fields and have $J^\mu J_\mu$ structure as follows:
\begin{eqnarray}
&&O_1^{cb}=[V_{ud}(\bar{d}_{\alpha}u_{\alpha})_{V-A}+V_{us}(\bar{s}_{\alpha}u_{\alpha})_{V-A}\
+V_{cd}(\bar{d}_{\alpha}c_{\alpha})_{V-A}+V_{cs}(\bar{s}_{\alpha}c_{\alpha})_{V-A}]
(\bar{c}_{\beta}b_{\beta})_{V-A}\,,\nonumber\\
&&O_2^{cb}=[V_{ud}(\bar{d}_{\alpha}u_{\beta})_{V-A}+V_{us}(\bar{s}_{\alpha}u_{\beta})_{V-A}\
+V_{cd}(\bar{d}_{\alpha}c_{\beta})_{V-A}+V_{cs}(\bar{s}_{\alpha}c_{\beta})_{V-A}]
(\bar{c}_{\beta}b_{\alpha})_{V-A}\,,\nonumber\\
&&O_3=(\bar q_\alpha b_\alpha)_{V-A}\sum\limits_{q'}(\bar q'_\beta
q'_\beta)_{V-A},\ \ O_4=(\bar q_\beta
b_\alpha)_{V-A}\sum\limits_{q'}(\bar q'_\alpha q'_\beta)_{V-A},
\nonumber\\
&&O_5=(\bar q_\alpha b_\alpha)_{V-A}\sum\limits_{q'}(\bar q'_\beta
q'_\beta)_{V+A},\ \ O_6=(\bar q_\beta
b_\alpha)_{V-A}\sum\limits_{q'}(\bar q'_\alpha q'_\beta)_{V+A},
\nonumber\\
&&O_7=\frac{3}{2}(\bar q_\alpha
b_\alpha)_{V-A}\sum\limits_{q'}e_{q'}(\bar q'_\beta
q'_\beta)_{V+A},\ \ O_8=\frac{3}{2}(\bar q_\beta
b_\alpha)_{V-A}\sum\limits_{q'}e_{q'}(\bar q'_\alpha
q'_\beta)_{V+A},
\nonumber\\
&&O_9=\frac{3}{2}(\bar q_\alpha
b_\alpha)_{V-A}\sum\limits_{q'}e_{q'}(\bar q'_\beta
q'_\beta)_{V-A},\ \ O_{10}=\frac{3}{2}(\bar q_\beta
b_\alpha)_{V-A}\sum\limits_{q'}e_{q'}(\bar q'_\alpha
q'_\beta)_{V-A},
\end{eqnarray}
 where
$(\bar{q}_1q_2)_{V-A}=\bar{q}_1{\gamma}^{\mu}(1-{\gamma}_5)q_2$.
The operators $O_1$ and $O_2$ are the current-current (tree)
operators, $O_3,...,O_6$ are the QCD-penguin operators and
$O_7,...,O_{10}$ are the electroweak penguin operators. Since we
calculate the decay up-to leading order, we just consider the
contribution of $O_1$ and $O_2$.

Here we apply the so-called naive factorization to $H_{eff}$ i.e.
the operators $O_i$ \cite{naive-fac}, so the nonleptonic two-body
decay amplitude $T$ can be reduced to a product of a transition
matrix element of a weak current $\langle M_1|J^{\mu}|B_c\rangle$
and an annihilation matrix element of another weak current $\langle
M_2|J_{\mu}|0\rangle$:
\begin{eqnarray}\label{non-am0}
T=\langle M_1M_2|H_{eff}|B_c\rangle\approx \langle
M_1|J^{\mu}|B_c\rangle\langle M_2|J_{\mu}|0\rangle\,,
\end{eqnarray}
while the annihilation matrix element is relating to a decay
constant directly. The reason why we adopt the naive factorization
here is that it works well enough due to the fact that all the
decays concerned in this paper are `constrained' to those in them
the quark $c$ as a `spectator' goes from initial $B_c$ meson into
the final meson $M_1$ always, thus as pointed by the authors of
\cite{naive,naive1}, in the concerned cases the corrections to the
naive factorization are suppressed.

Since $M_1=\chi_c(h_c)$, the matrix element $\langle
M_1|J^{\mu}|B_c\rangle$ is just the hadronic weak-current matrix
element appearing in the previous subsection, but different from it
by momentum transfer being fixed (owing to the decays are of one to
two-body). The annihilation matrix element $\langle
M_2|J_{\mu}|0\rangle$ with $J^{\mu}=(\bar{q}_1q_2)_{V-A}$ is related
to the decay constant of a `common meson' $M_2$ and can be measured
via proper processes generally.

Precisely, let us now `restrict ourselves' to analyze the $B_c$
nonleptonic decays to the $P$-wave charmonium and the ${\pi}^+$,
${\rho}^+$, etc, which are governed by the weak decay
$\bar{b}\to\bar{c}u\bar{d}$, or to the $P$-wave charmonium and
$K^+$, $K^*$, etc, which are governed by the weak decay
$\bar{b}\to\bar{c}u\bar{s}$. As an example, under naive
factorization, we have the decay amplitude of $B_c\to
{\chi}_{c0}\rho^+$ as follows:
\begin{eqnarray}\label{non-am1}
T(B_c \to{\chi}_{c0}{\rho}^+)=\frac{G_F}{\sqrt2}V_{cb}V^*_{ud}
a_1(\mu)\langle{\chi}_{c0}|J^{\mu}|B_c\rangle\langle{\rho}^
+|J_{\mu}|0\rangle\,,
\end{eqnarray}
here $a_1=c_1+\frac{1}{N_c}c_2$ and $N_c=3$ is the number of
colors.

Since $\langle M_2|J_{\mu}|0\rangle$ is relating to the decay
constant of the meson $M_2$ directly, so to calculate the widths
of the non-leptonic decays is straightforward when the
weak-current transition matrix elements $\langle
M_1|J^{\mu}|B_c(P)\rangle$ are well calculated. Thus one may see
that the problem to calculate the non-leptonic decays is
essentially attributed to calculating the hadronic weak-current
matrix elements $\langle M_1|V^{\mu}|B_c(P)\rangle$ and $\langle
M_1|A^{\mu}|B_c(P)\rangle$ appearing in the above subsection for
semileptonic decays.

\section{Computation of the transition-matrix elements for weak-currents}

From the section above, we can see that to calculate the weak
currents matrix elements $\langle M_1|J^{\mu}|B_c(P)\rangle$ is the
key problem for the concerned semileptonic and nonleptonic decays,
so let us now explain the reason why and show how to apply the newly
developed method \cite{improv} to calculate the matrix elements. In
fact it is also to prepare necessary formulae for final numerical
calculations.

Here the weak-current matrix elements are for `transitions' from a
state of a double heavy meson to another double heavy meson. Due
to the mass difference of the two states, the relativistic effects
for the transitions are great, that a proper formulation to deal
with the relativistic effects is desired. It is known that the
approach of relativistic B-S equation for the bound states and
Mandelstam formulation for the transition matrix elements may be
taken into account quite well, and furthermore the B-S equation
and Mandelstam formulation even under `instantaneous
approximation' still works, because here the involved mesons are
double heavy. While the newly developed method \cite{improv},
which applies the `instantaneous approximation' to the current
matrix elements and B-S equation completely, should be better than
the original one in Ref.~\cite{previous}, where the `instantaneous
approximation' is applied incompletely. The 'completeness' here
means to apply it to the B-S equation, the solutions (B-S wave
functions) and the transition matrix element (under Mandelstam
formulation) properly, and let us outline it below.

According to the Mandelstam formulation \cite{Mand}, the
corresponding hadronic matrix elements of weak current between the
double heavy meson $B_c$ in initial state and the double heavy
meson $\chi_c(h_c)$ in final state, appearing in
Eq.~(\ref{weak-current}), Eq.~(\ref{non-am0}) and
Eq.~(\ref{non-am1}), can be written as:
\begin{eqnarray}\label{AA}
&&\langle \chi_c(h_c)(P_f)|J^{\mu}|B_c(P)\rangle\nonumber\\
&&\displaystyle=i\int\frac{d^{4}qd^4q'}{(2\pi)^{4}}
Tr\left[\overline\chi_{\chi_c(h_c)}(P',q')
(\not\!{p_{1}}-m_{1})\chi_{_{B_c}}(P,q)V_{cb}\gamma^{\mu}
(1-\gamma_5)\delta(p_1-p'_1)\right]\nonumber\\
&&\displaystyle=i\int\frac{d^{4}q}{(2\pi)^{4}}Tr
\left[\overline\chi_{\chi_c(h_c)}(P',q')
(\alpha_1\not\!{P}+\not\!q-m_{1})\chi_{_{B_c}}(P,q)V_{cb}
\gamma^{\mu}(1-\gamma_5)\right],
\end{eqnarray}
where $p_1 = \alpha_1P + q$ ($\alpha_1 \equiv \frac{m_1}{ m_1 +
m_2}$), $p_2 = \alpha_2P-q$ ($\alpha_2\equiv\frac{ m_2}{ m_1 +
m_2}$) are the momenta of $c$-quark and $\bar{b}$-quark respectively
inside $B_c$ meson; $p'_1 = \alpha'_1P_f + q'$ ($\alpha'_1 \equiv
\frac{m'_1}{ m'_1 + m'_2}$), $p'_2 = \alpha'_2P_f-q'$
($\alpha'_2\equiv\frac{ m'_2}{ m'_1 + m'_2}$) are the momenta of
$c$-quark and $\bar{c}$-quark respectively inside the $P$-wave
charmonium  $\chi_c(h_c)$; moreover, for the final result (the last
line of Eq.~(\ref{AA})) we have $P=P_f+p_l+p_\nu$ and
$q'=\alpha_1P+q-\alpha'_1P_f$.

The newly developed method \cite{improv} essentially is to apply the
`instantaneous approximation' to the current matrix elements and the
B-S equation completely, to outline it and for `applying the
instantaneous approximation' in a covariant way, we need to
decompose the relative momentum $q$ into two components: the
time-like one $q^\mu_\parallel$ and the space-like one $q^\mu_\perp$
as follows:
$$q^\mu=q^\mu_\parallel+q^\mu_\perp,\ \ q^\mu_\parallel\equiv
\frac{ P\cdot q}{M^2}P^\mu,\ \ q^\mu_\perp\equiv
q^\mu-q^\mu_\parallel,$$
$$P'{^{\mu}}=P'{^{\mu}}_{\parallel}+P'{^{\mu}}_{\perp}\,,\;\;\;\;
P'{^{\mu}}_{\parallel}\equiv \frac{(P\cdot
P')}{M^{2}}P^{\mu}\,,\;\;\;\;P'{^{\mu}}_{\perp}\equiv
P'{^{\mu}}-P'{^{\mu}}_{\parallel}\,;$$ and
$$q'^{\mu}=q'^{\mu}_{\parallel}+q'^{\mu}_{\perp},\,\;\;\;\;
q'^{\mu}_{\parallel}\equiv (P\cdot q'/M^{2})P^{\mu},\,\;\;\;\;
q'^{\mu}_{\perp}\equiv q'^{\mu}-q'^{\mu}_{\parallel},\,$$ where
$M$ is the mass of the meson $B_c$, and we may further have two
Lorentz invariant variables $q_P\equiv \frac{P\cdot q}{M}$ and
$q_T\equiv \sqrt{-q_\perp^2}$.

The `instantaneous approximation' applying to the matrix element is
just to carry out the integration of $dq^\mu_\parallel$ by a contour
one on Eq.~(\ref{AA}) precisely and to obtain the result below:
\begin{eqnarray}\label{AA1}
&\displaystyle\langle
\chi_c(h_c)(P_f)|J^{\mu}|B_c(P)\rangle=
i\int\frac{d^{4}q}{(2\pi)^{4}}Tr\left[\overline\chi_{\chi_c(h_c)}(P',q')
(\alpha_1\not\!{P}+\not\!q-m_{1})
\chi_{_{B_c}}(P,q)V_{cb}\gamma^{\mu}(1-\gamma_5)\right] \nonumber\\
&\displaystyle=\int\frac{d^3q_{\bot}}{(2\pi)^3}Tr
\Bigg\{\Big[\bar{\varphi}^{\prime++}(q_{\bot}^{\prime})\frac{\not\!P}{M}{\varphi}^{++}
(q_{\bot})+\bar{\varphi}^{\prime++}(q_{\bot}^{\prime})\frac{\not\!P}{M}{\psi}^{+-}
(q_{\bot})\nonumber\\
&\displaystyle-\bar{\psi}^{\prime-+}(q_{\bot}^{\prime})\frac{\not\!P}{M}{\varphi}^{++}
(q_{\bot})-\bar{\psi}^{\prime+-}(q_{\bot}^{\prime})\frac{\not\!P}{M}{\varphi}^{--}
(q_{\bot})\nonumber\\
&\displaystyle+\bar{\varphi}^{\prime--}(q_{\bot}^{\prime})\frac{\not\!P}{M}{\psi}^{-+}
(q_{\bot})-\bar{\varphi}^{\prime--}(q_{\bot}^{\prime})\frac{\not\!P}{M}{\varphi}^{--}
(q_{\bot})\Big]{\gamma}^{\mu}(1-{\gamma}_5)\Bigg\},
\end{eqnarray}
where:
\begin{eqnarray}\label{Salp+-}
&\displaystyle\varphi^{++}(q_\perp)=\frac{\Lambda_{1}^{+}(q_\perp)\eta(q_\perp)
\Lambda_{2}^{+}(q_\perp)}{M-\omega_{1}-\omega_{2}}\,,\;\;\;
&\bar{\varphi}^{\prime++}(q'_{P\perp})=\frac{\Lambda_{2}^{\prime+}(q'_{P\perp})\bar{\eta}(q'_{P\perp})
\Lambda_{1}^{\prime+}(q'_{P\perp})}{E_f-\omega'_{1}-\omega'_{2}}\,,\nonumber\\
&\displaystyle\varphi^{--}(q_\perp)=-\frac{\Lambda_{1}^{-}(q_\perp)\eta(q_\perp)
\Lambda_{2}^{-}(q_\perp)}{M+\omega_{1}+\omega_{2}}\,,\;\;\;
&\bar{\varphi}^{\prime--}(q'_{P\perp})=-\frac{\Lambda_{2}^{\prime-}(q'_{P\perp})\bar{\eta}(q'_{P\perp})
\Lambda_{1}^{\prime-}(q'_{P\perp})}{E_f+\omega'_{1}+\omega'_{2}}\,,\nonumber\\
&\displaystyle{\psi}^{-+}(q_\perp)= \frac{
\Lambda^{-}_{1}(q_{P\perp}) \eta(q_\perp)\Lambda^{+}_{2}(q_{P\perp})
}{M-\omega_{2}-\omega'_{2}-E_f}\,,
&\bar{\psi}^{\prime-+}(q'_{P\perp})=
\frac{\Lambda'^{-}_2(q'_{P\perp})
\bar{\eta}'(q'_{P\perp})\Lambda'^{+}_1(q'_{P\perp})}
{M-\omega_{2}-\omega'_{2}-E_f}\,,\nonumber \\
&\displaystyle{\psi}^{+-}(q_\perp)=
\frac{\Lambda^{+}_{1}(q_{P\perp})\eta(q_\perp)\Lambda^{-}_{2}(q_{P\perp})
}{M+\omega_{2}+\omega'_{2}-E_f}\,,
&\bar{\psi}^{\prime+-}(q'_{P\perp})=
\frac{\Lambda'^{+}_2(q'_{P\perp})
\bar{\eta'}(q'_{P\perp})\Lambda'^{-}_1(q'_{P\perp})
}{M+\omega_{2}+\omega'_{2}-E_f}\,,
\end{eqnarray}
${\varphi}^{ij}(q_{\bot}),\,\psi^{ij}(q_{\bot})$ and
${\bar{\varphi}}^{\prime{ij}}(q_{P\bot}^{\prime}),\,\bar{\psi}^{\prime{ij}}(q_{P\bot}^{\prime})$
are B-S wave functions as the B-S equation solutions under
`complete instantaneous approximation' \cite{chenjk,spectr} and
with `energy projection' $\Lambda^\pm$ of the mesons in initial
and finial states properly. The precise definitions of the `energy
projection' and the B-S `vertex' ${\eta}_{P}$, $\bar{\eta}_{P}$
(${\eta'}_{P}$, $\bar{\eta'}_{P}$) are presented in Appendix B.
One may also see that the four equations,
Eqs.~(\ref{Salp10},~\ref{Salp11},~\ref{Salp12}), are B-S equations
under the complete instantaneous approximation, instead of the
incomplete instantaneous approximation which only considering the
Eq.~(\ref{Salp10}).

Namely the `improvements' from the `newly development method' are
attributed to: i). with the complete instantaneous approximation to
current matrix element, as a result, there are six terms in the
squared bracket of Eq.~(\ref{AA1}) instead of the first term
\begin{eqnarray}\label{previous}
\langle \chi_c(h_c)(P_f)|J^{\mu}|B_c(P)\rangle
=\int\frac{d^3q_{\bot}}{(2\pi)^3}Tr
\Big\{\bar{\varphi}^{\prime++}(q_{\bot}^{\prime})\frac{P}{M}{\varphi}^{++}
(q_{\bot}){\gamma}^{\mu}(1-{\gamma}_5)\Big\}\,
\end{eqnarray} is only kept;
ii). the B-S wave functions hidden in
${\varphi}^{ij}(q_{\bot}),\,\psi^{ij}(q_{\bot})$ and
${\bar{\varphi}}^{\prime{ij}}(q_{P\bot}^{\prime}),\,
\bar{\psi}^{\prime{ij}}(q_{P\bot}^{\prime})$ are solved under
complete instantaneous approximation to the B-S equation. For the
point i), since the considered double heavy meson, $B_c$ or
$\chi_c(h_c)$, is weak binding system i.e. the binding energy
$\varepsilon \equiv M-\omega_1-\omega_2$ (or $\varepsilon \equiv
E_f-\omega'_1-\omega'_2$) is small ($\frac{\varepsilon}{M}\ll
O(1)$), thus from  Eq.~(\ref{Salp+-}) we are sure that
${\varphi}^{++}(q_{\bot})$ and
${\bar{\varphi}}^{\prime++}(q_{P\bot}^{\prime})$ are much greater
than the others ${\varphi}^{ij}(q_{\bot}),\,\psi^{ij}(q_{\bot})$
and ${\bar{\varphi}}^{\prime{ij}}(q_{P\bot}^{\prime}),
\,\bar{\psi}^{\prime{ij}}(q_{P\bot}^{\prime})$, so that using the
Eq.~(\ref{previous}) instead of Eq.~(\ref{AA1}) is a very good
approximation, which we have precisely examined by considering the
decay $B_c\to {\chi}_{c0}l{\nu}_l$ as an example: in fact, the
contributions of the second term and third term of Eq.~(\ref{AA1})
to the form factor are less than the one of first term of
Eq.~(\ref{AA1}) roughly by a factor $10^{-2}\sim10^{-3}$ times. If
the first three terms are considered, the decay width is
$1.85\times10^{-15}$ GeV, while if only the first term is
considered, the decay width is $1.87\times10^{-15}$ GeV, i.e. the
two results are very similar. So the approximation is very good
and we may use Eq.~(\ref{previous}) instead of Eq.~(\ref{AA1}) to
compute the weak-current matrix elements safely.

\section{Numerical calculations and results with proper comparisons}

In this section, based on the formulations obtained in the paper,
we evaluate the decay widths for semileptonic and nonleptonic
decays and some interesting quantities else for semileptonic
decays, such as form factors and charged lepton spectrum etc and
then discuss them briefly.

First of all, we need to fix the parameters appearing in the
framework. We adjusted the parameters $a=e=2.7183$, $\lambda=0.21$
GeV$^2$, ${\Lambda}_{QCD}=0.27$ GeV, $m_b=4.96$ GeV, $m_c=1.62$
GeV and $V_0$ for the B-S kernel as those in Refs.
\cite{spectr,glwang,wanggl}, which as the best input for
spectroscopy, then the spectra of the mesons and the masses
$M_{B_c}=6.276$ GeV, $M_{{\chi}_{c0}}=3.414$ GeV,
$M_{{\chi}_{c1}}=3.510$ GeV, $M_{{\chi}_{c2}}=3.555$ GeV,
$M_{{h}_{c}}=3.526$ GeV etc \cite{spectr}, which are used in this
paper, are obtained, moreover the decay constants, average
energies as well as annihilations of quarkonia are fitted
\cite{wanggl,glwang,glwang2}.

With the obtained B-S wave functions (under the formulation defined
in Appendix B) and as a next step, we substitute the functions into
${\varphi}^{++}(q_{\bot})$ and
${\bar{\varphi}}^{\prime++}(q_{P\bot}^{\prime})$, so that they are
related to the components of the B-S wave functions precisely as
depicted in Appendix C. With the formula Eq.~(\ref{previous}),
finally we represent the hadronic transition weak-current matrix
elements as proper integrations of the components of the B.-S. wave
functions. As final results of this paper, the decay widths for the
semileptonic and nonleptonic decays and some interesting quantities
else for the semileptonic decays, such as form factors and charged
lepton spectrum etc, are straightforwardly calculated numerically.
In the following subsections we present the results for the
semileptonic decays and nonleptonic decays separately.

\subsection{The semi-leptonic decays}

\begin{table}\caption{The semileptonic decay widths (in the unit $10^{-15}$GeV)}
\begin{center}
\begin{tabular}{|c|c|c|c|c|c|c|c|}
\hline\hline
Mode& This work &\cite{Bra}&\cite{Bww}&\cite{wqd}&\cite{qwerr}&\cite{Az}&\cite{Lu} \\
 \hline
$B_c^+\to{\chi}_{c0}e\nu$&$1.87\pm0.46$&1.27&2.52&1.55&1.69&2.60$\pm$0.73&\\

\hline
$B_c^+\to{\chi}_{c0}\tau\nu$&$0.23\pm0.12$&0.11&0.26&0.19&0.25&0.7$\pm$0.23&\\
\hline

$B_c^+\to{\chi}_{c1}e\nu$&$1.52\pm0.45$&1.18&1.40&0.94&2.21&2.09$\pm$0.60&\\
\hline
$B_c^+\to{\chi}_{c1}\tau\nu$&$0.14\pm0.10$&0.13&0.17&0.10&0.35&0.21$\pm$0.06&\\
\hline $B_c^+\to{\chi}_{c2}e\nu$&$1.50\pm0.39$&2.27&2.92&1.89&2.73& &\\
\hline$B_c^+\to{\chi}_{c2}\tau\nu$&$0.12\pm0.07$&0.13&0.20&0.13&0.42& &\\
\hline$B_c^+\to
h_{c}e\nu$&$3.98\pm1.10$&1.38&4.42&2.4&2.51&2.03$\pm$0.57&4.2$\pm$2.1\\
\hline$B_c^+\to
h_{c}\tau\nu$&$0.28\pm0.20$&0.11&0.38&0.21&0.36&0.20$\pm$0.05&0.53$\pm$0.26\\
\hline\hline
\end{tabular}
\end{center}
\end{table}

\begin{figure}
\centering
\includegraphics[height=7cm]{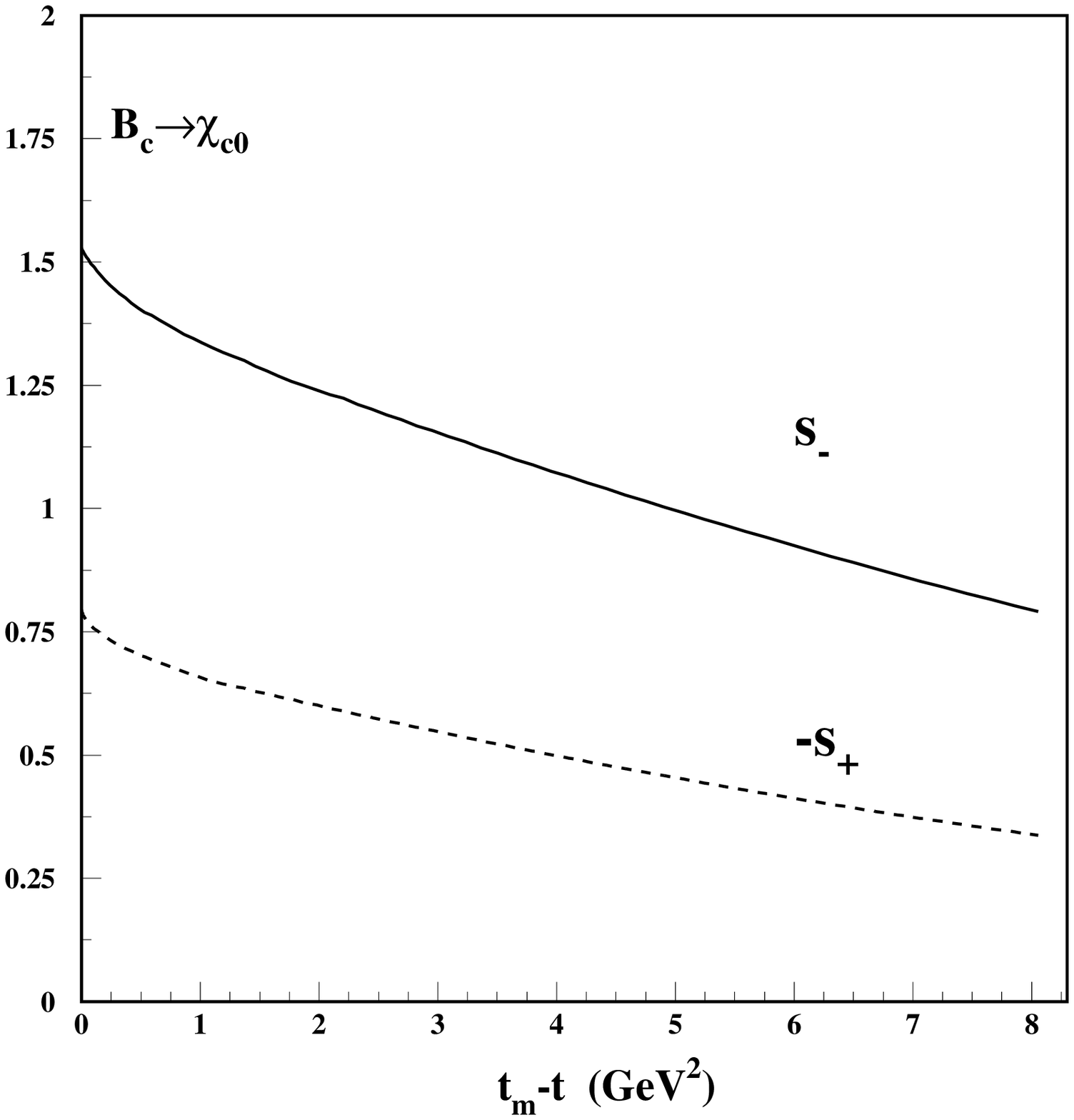}
\includegraphics[height=7cm]{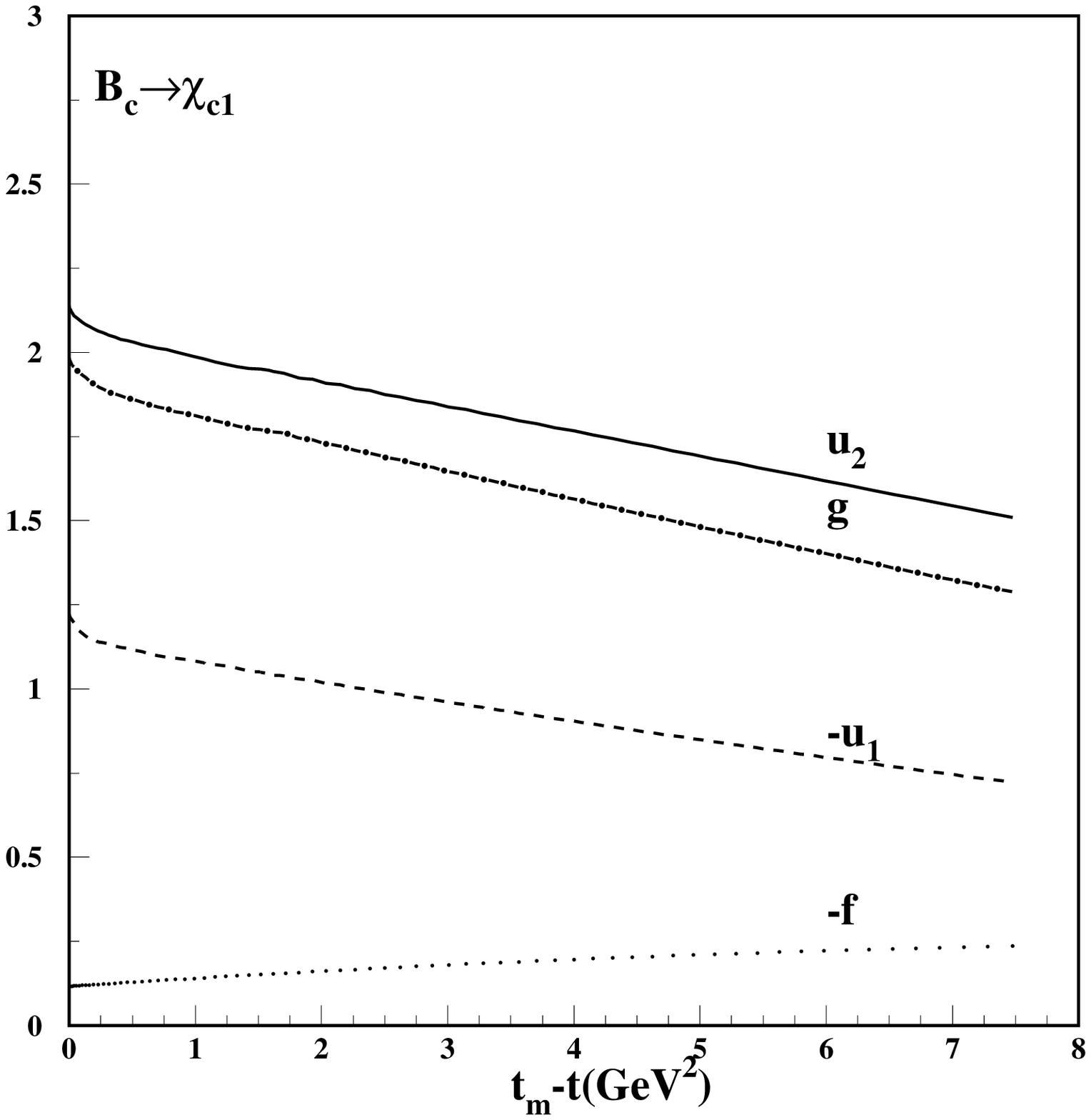}
\includegraphics[height=7cm]{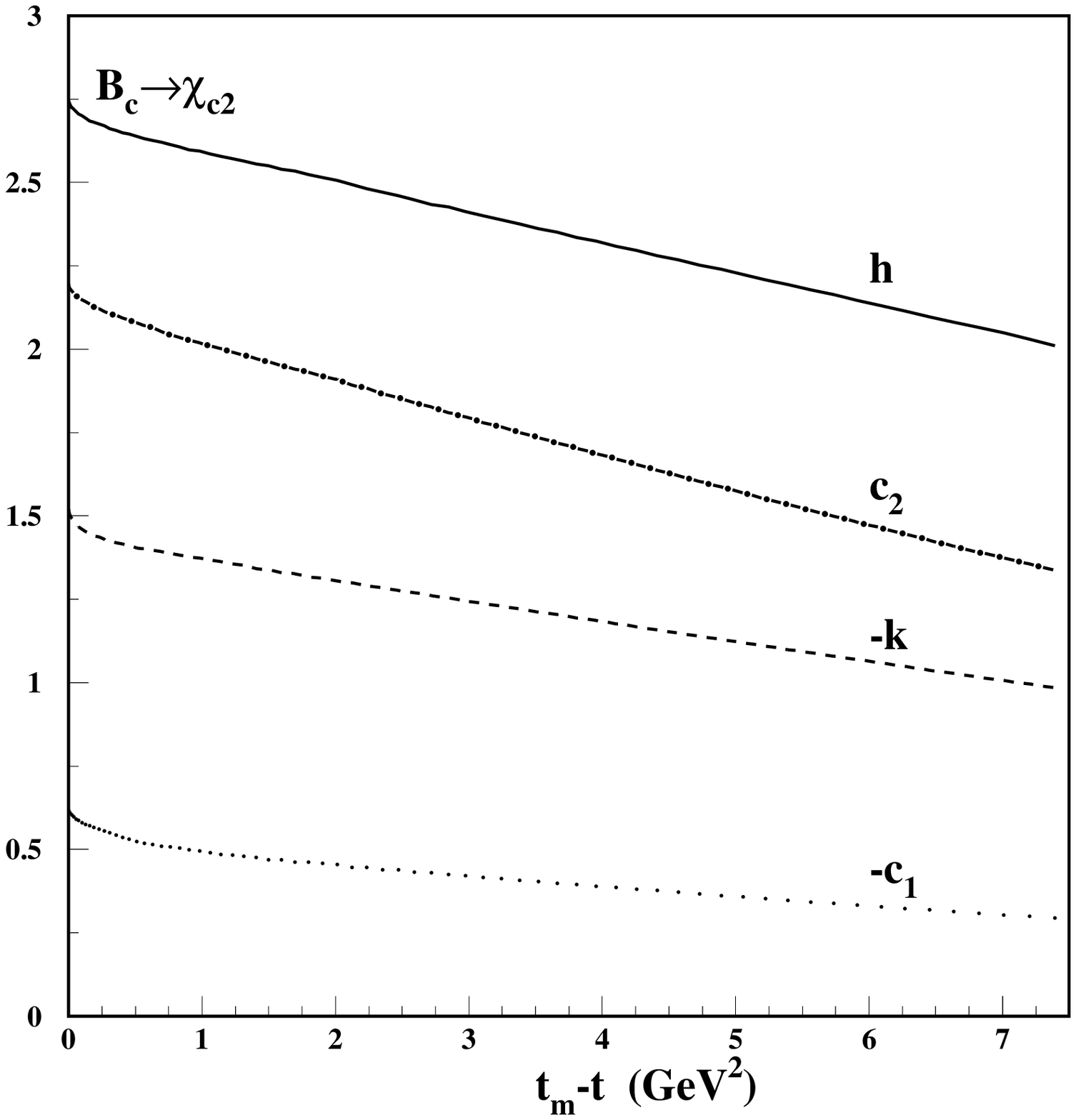}
\includegraphics[height=7cm]{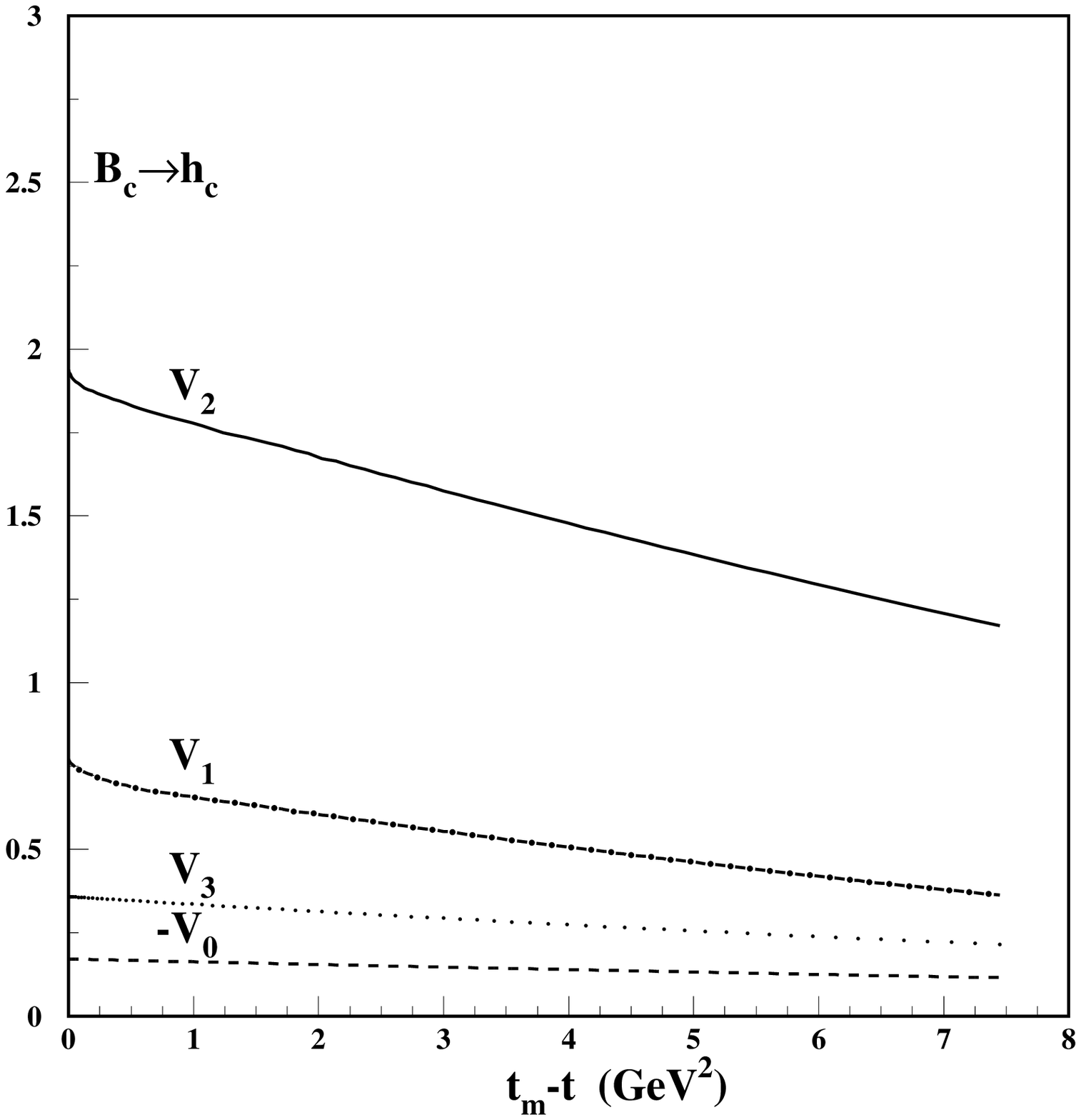}
\caption{The form factors of the $B_c$ decays to a $P$-wave
charmonium defined as in Eq.(\ref{f-factor1}), Eq.(\ref{f-factor2}),
Eq.(\ref{f-factor3}) and Eq.(\ref{f-factor4}) and
$t=q^2=(P-P_f)^2=M^2+M_f^2-2ME_f$ ($t_m$ is the maximum of $t$).}
\end{figure}

\begin{figure}
\centering
\includegraphics[height=7cm]{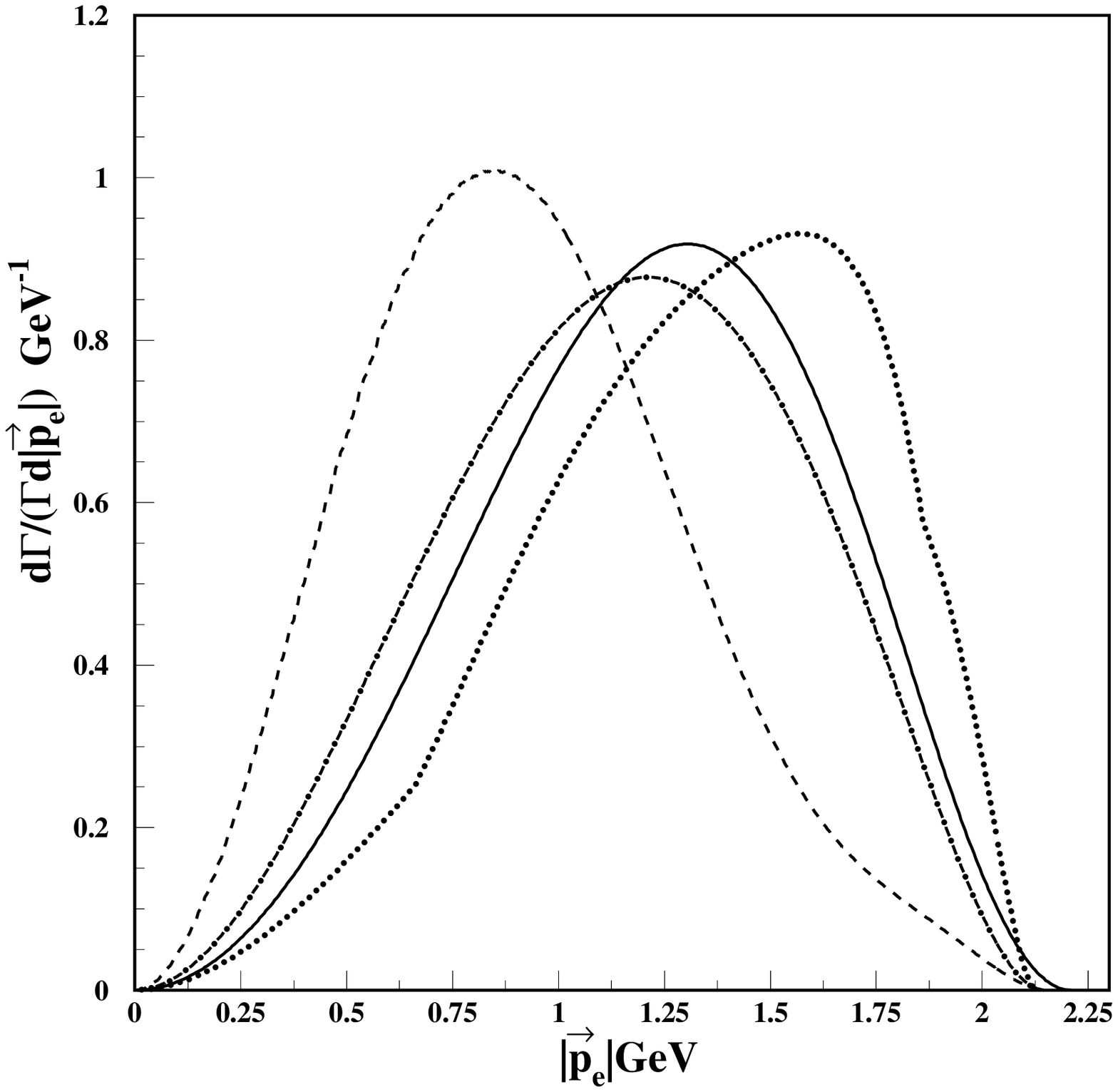}
\includegraphics[height=7cm]{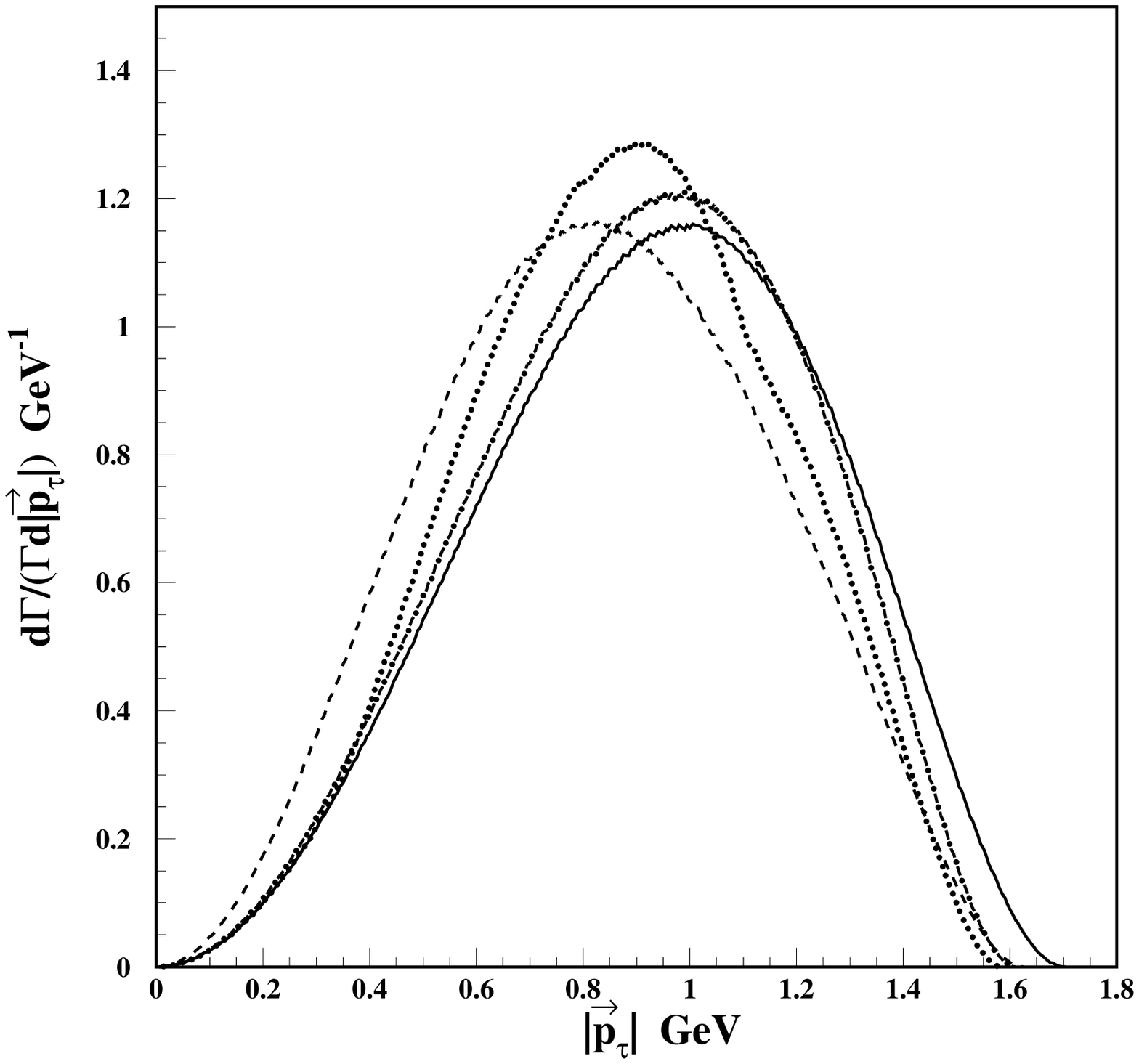}
\caption{The energy spectrums of the charged lepton in the $B_c$
semileptonic decays to $P$-wave charmoniums. The left figure is for
$B_c \to \chi_{c0,1,2}(h_c)e\nu$ and the right figure is for $B_c
\to \chi_{c0,1,2}(h_c)\tau\nu$. Where the solid lines are the
results for ${\chi}_{c0}$, the dash lines are for ${\chi}_{c1}$, the
dot lines are for ${\chi}_{c2}$ and the dot-dash lines are for
$h_c$.}
\end{figure}

\begin{figure}
\centering
\includegraphics[height=7cm]{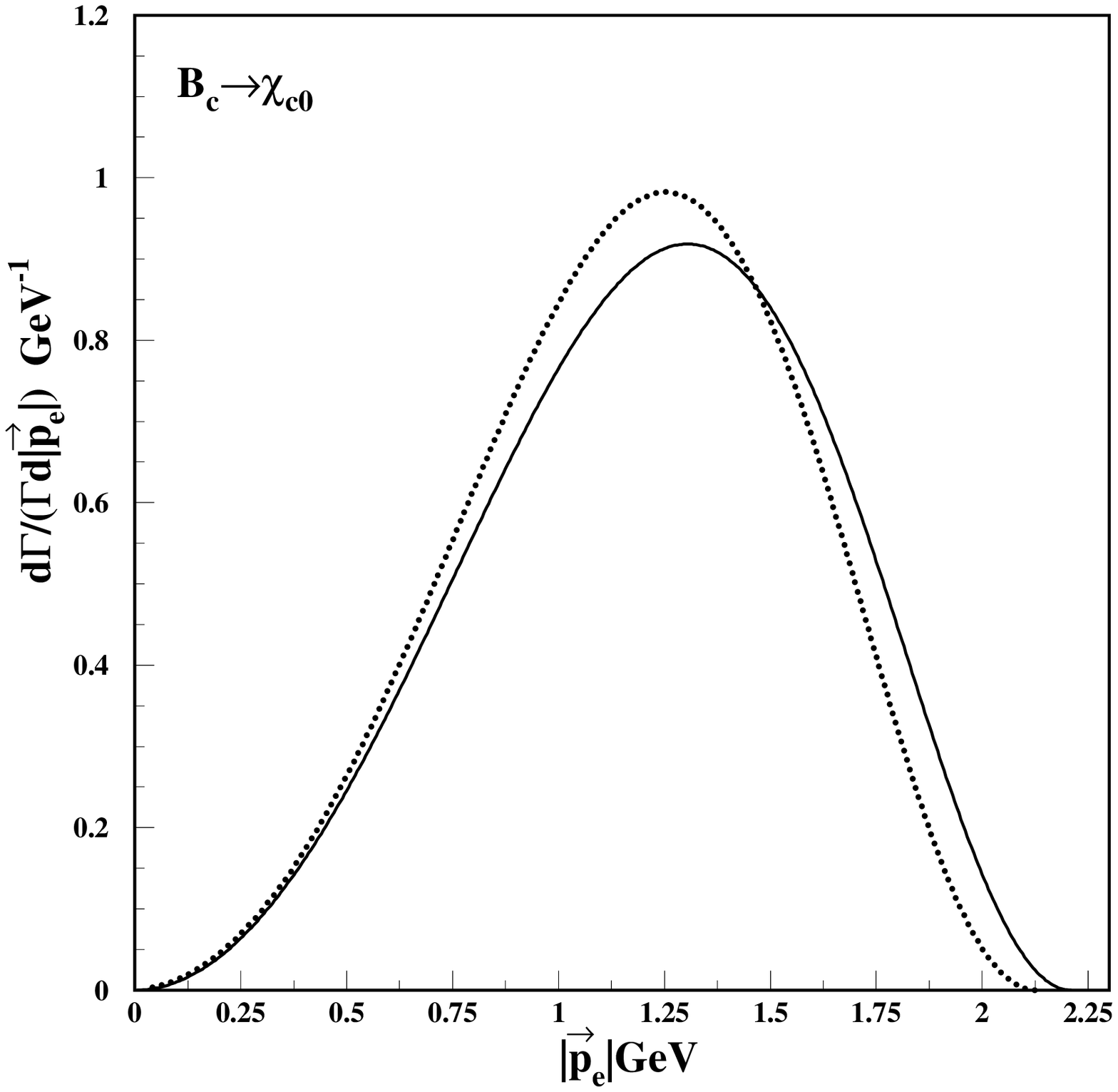}
\includegraphics[height=7cm]{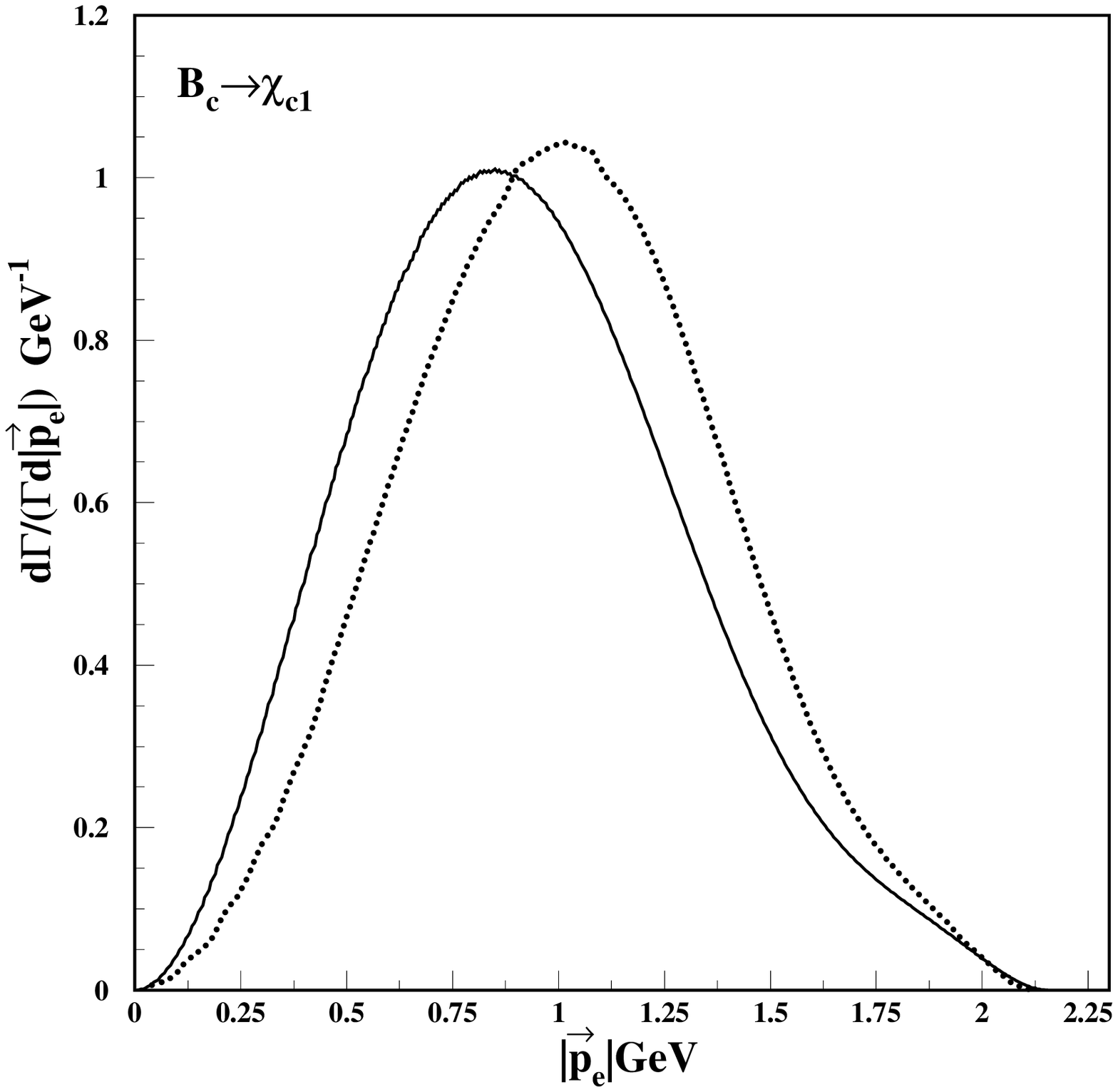}
\includegraphics[height=7cm]{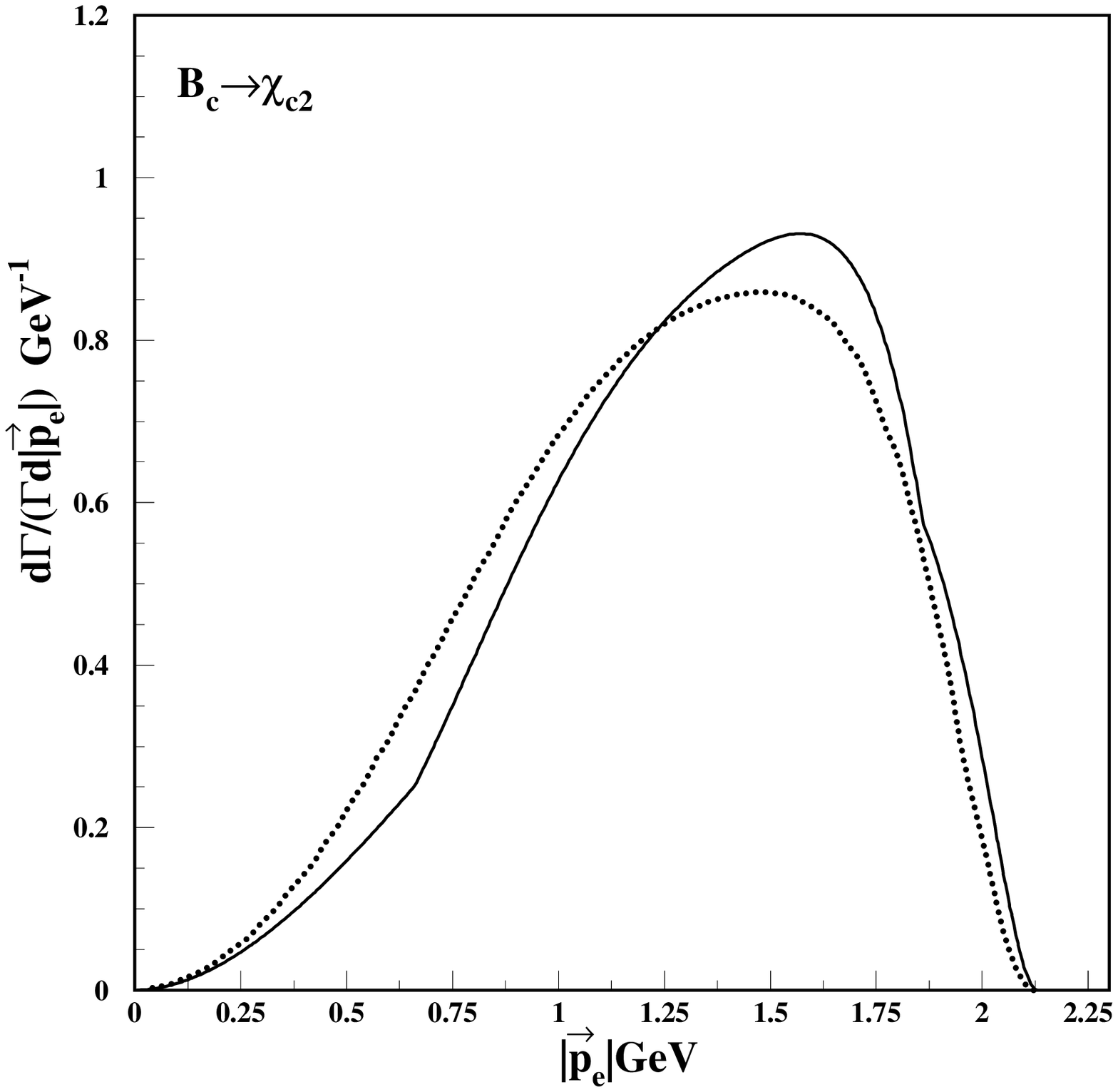}
\includegraphics[height=7cm]{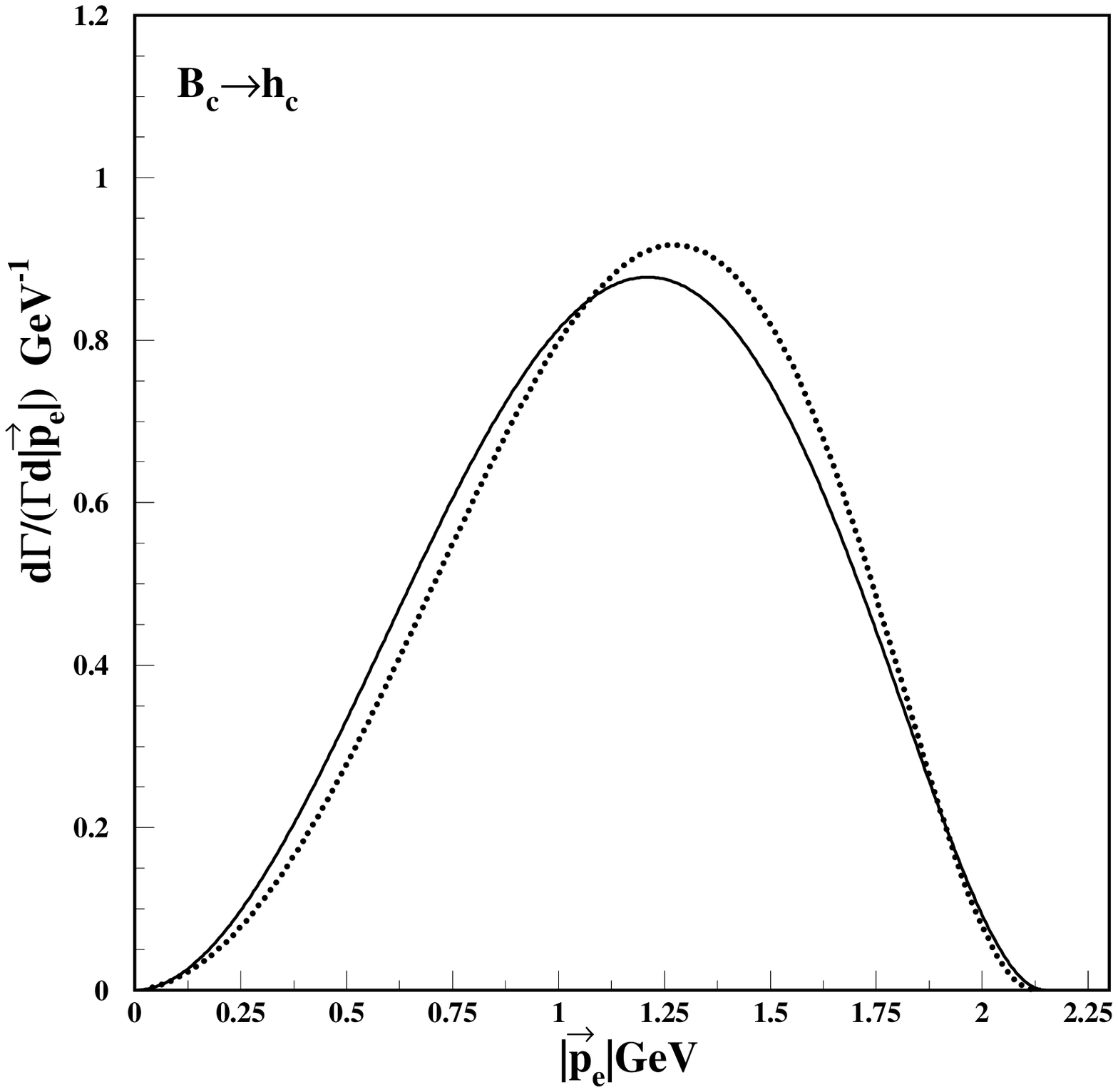}
\caption{The energy spectrums of the charged lepton in the $B_c$
semileptonic decays to $P$-wave charmoniums respectively. The solid
lines are the results of this work, the dash lines are the results
of \cite{qwerr}.}
\end{figure}

When the weak current transition matrix element for a definite
semi-leptonic decay is calculated precisely and the values of the
CKM matrix elements $|V_{ud}|=0.974$, $|V_{us}|=0.225$,
$|V_{bc}|=0.0406$ \cite{pdg} are given, not only the decay width
can be calculated straightforwardly, but also the form factors may
be extracted out. Moreover as `semifinished product', the spectrum
of the charged lepton which may be measurable experimentally can
be also acquired too. Namely the functions $\alpha$, $\beta_{++}$,
$\beta_{+-}$, $\beta_{-+}$, $\beta_{--}$, $\gamma$ appearing in
the spectrum of the charged lepton (see Eq.~(\ref{differ})) are
related to the form factors directly as shown in Appendix A
precisely. Therefore when we calculate and present the results for
semi-leptonic decays, not only those of the decay widths but also
the spectrums of the charged lepton in the decays are considered.
Since $\tau$ lepton is quite massive and $m_\mu\simeq m_e$ is
quite a good approximation for the $B_c$ meson decays, so when we
calculate and present the widths and the spectrums of the charged
lepton for the decays, only the cases that the lepton being
electron or $\tau$ are considered.

Note that since the input B-S wave functions by solving the B-S
equation for the double heavy mesons which are involved in the
transition matrix elements of weak current have uncertainties, due
to the parameters fitting to fix the B-S kernel and quark masses,
the way to solve the B-S equation numerically, and the
approximation from Eq.~(\ref{AA1}) to Eq.~(\ref{previous}) for the
transition matrix elements of the weak currents is taken etc, so
in the numerical results obtained finally there are certain
errors. To consider the uncertainties caused by the input
parameters, we changed all the input parameters simultaneously
within $5\%$ of the center values, then we get the uncertainties
of numerical results for the semi-leptonic decays and the
non-leptonic decays shown in Table.~I. We find that the
uncertainties of the decays $B_c\to h_c(\chi_{c})+e+\nu_{e}$ vary
up to $30\%$ of center values, while the uncertainties of $B_c\to
h_c(\chi_{c})+\tau+\nu_{\tau}$ are up to $60\%$ in Table.~I, the
reason is that the phase spaces for $B_c\to
h_c(\chi_{c})+\tau+\nu_{\tau}$ are smaller than the ones for
$B_c\to h_c(\chi_{c})+e+\nu_{e}$ because of the heavy $\tau$
lepton, and the the uncertainties for the former are more
sensitive to the changes of the phase space than the latter.

To compare with the results obtained by the other approaches, we
present the decay widths calculated out this work with error bar
and the results obtained by the other approaches by putting them
together in a table i.e. Table~I.

In addition we also present the obtained form factors and the
spectrums of the charged lepton in the decays in Fig.~3 and Fig.~4
respectively. To compare with the results of the previous work
Ref.~\cite{qwerr}, we draw the curves of the spectrums of charged
lepton obtained by this work and the work Ref.~\cite{qwerr} in
Fig.~5. Whereas in order to see the tendency of the form factors
and the lepton spectrum clearly and we suspect that at present
stage it is enough, so in the figures we draw the curves with the
center values but not involve the errors precisely.

\subsection{The non-leptonic decays}

\begin{table}\caption{The nonleptonic decay widths (in the unit $10^{-15}$GeV)}
\begin{center}
\begin{tabular}{|c|c|c|c|c|c|}
\hline\hline
Mode& This work &\cite{Bra}&\cite{Bww}&\cite{wqd}&\cite{qwerr}\\
\hline
$B_c^+\to{\chi}_{c0}{\pi}^+$&$(0.34\pm0.04)a_1^2$&0.23$a_1^2$&0.622$a_1^2$&0.28$a_1^2$&0.317$a_1^2$\\
$B_c^+\to{\chi}_{c1}{\pi}^+$&$(0.023\pm0.002)a_1^2$&0.22$a_1^2$&0.076$a_1^2$&0.0015$a_1^2$&0.0815$a_1^2$\\
$B_c^+\to{\chi}_{c2}{\pi}^+$&$(0.24\pm0.05)a_1^2$&0.41$a_1^2$&0.518$a_1^2$&0.24$a_1^2$&0.277$a_1^2$\\
 $B_c^+\to h_c{{\pi}^+}$&$(1.10\pm0.16)a_1^2$&0.51$a_1^2$&1.24$a_1^2$&0.58$a_1^2$&0.569$a_1^2$\\
\hline
$B_c^+\to{\chi}_{c0}{\rho}^+$&$(0.85\pm0.10)a_1^2$&0.64$a_1^2$&1.47$a_1^2$&0.73$a_1^2$&0.806$a_1^2$\\
$B_c^+\to{\chi}_{c1}{\rho}^+$&$(0.25\pm0.02)a_1^2$&0.16$a_1^2$&0.326$a_1^2$&0.11$a_1^2$&0.331$a_1^2$\\
$B_c^+\to{\chi}_{c2}{\rho}^+$&$(0.62\pm0.12)a_1^2$&1.18$a_1^2$&1.33$a_1^2$&0.71$a_1^2$&0.579$a_1^2$\\
$B_c^+\to h_c{{\rho}^+}$&$(2.50\pm0.50)a_1^2$&1.11$a_1^2$&2.78$a_1^2$&1.41$a_1^2$&1.40$a_1^2$\\
\hline
$B_c^+\to{\chi}_{c0}{K}^+$&$(0.026\pm0.003)a_1^2$&0.018$a_1^2$&0.0472$a_1^2$&0.022$a_1^2$&0.00235$a_1^2$\\
$B_c^+\to{\chi}_{c1}{K}^+$&$(0.0018\pm0.0002)a_1^2$&0.016$a_1^2$&0.0057$a_1^2$&0.00012$a_1^2$&0.0058$a_1^2$\\
$B_c^+\to{\chi}_{c2}{K}^+$&$(0.018\pm0.003)a_1^2$&0.031$a_1^2$&0.0384$a_1^2$&0.018$a_1^2$&0.00199$a_1^2$\\
 $B_c^+\to h_c{{K}^+}$&$(0.082\pm0.012)a_1^2$&0.039$a_1^2$&0.0939$a_1^2$&0.045$a_1^2$&0.0043$a_1^2$\\
\hline
$B_c^+\to{\chi}_{c0}{K}^{*+}$&$(0.050\pm0.006)a_1^2$&0.045$a_1^2$&0.0787$a_1^2$&0.041$a_1^2$&0.00443$a_1^2$\\
$B_c^+\to{\chi}_{c1}{K}^{*+}$&$(0.018\pm0.001)a_1^2$&0.01$a_1^2$&0.0201$a_1^2$&0.008$a_1^2$&0.00205$a_1^2$\\
$B_c^+\to{\chi}_{c2}{K}^{*+}$&$(0.037\pm0.007)a_1^2$&0.082$a_1^2$&0.0732$a_1^2$&0.041$a_1^2$&0.00348$a_1^2$\\
 $B_c^+\to h_c{{K}^{*+}}$&$(0.14\pm0.02)a_1^2$&0.077$a_1^2$&0.146$a_1^2$&0.078$a_1^2$&0.0076$a_1^2$\\
\hline \hline
\end{tabular}
\end{center}
\end{table}

The exclusive non-leptonic decays are of two-body in final states,
thus the hadronic transition matrix elements of weak-currents
appearing in Eq.~(\ref{non-am1}) have a fixed momentum transfer
$t=m^2_{M_2}$ (the mass squared of the other meson $M_2$ in the
decay $B_c\to M_1M_2$ and $M_1=\chi_c$ or $h_c$). In fact the
transition matrix elements have been already calculated in the
above subsection of semi-leptonic decays. To calculate the decay
widths, from Eq.~(\ref{non-am1}), now we need to calculate the
annihilation matrix element of the weak current such as $\langle
M_2|J_{\mu}|0\rangle $ additionally. It is known that the
annihilation matrix element is related to the `decay constant'
$f_{M_2}$ directly, and the decay constant $f_P$, $f_V$ or $f_A$
of a pseudoscalar meson, a vector meson or an axial vector meson
may be extracted from experimental data for the pure leptonic
decays of the relevant mesons, but they may also be calculated by
models, such as the one in Ref.~\cite{wanggl} although there are
some debates. In this work we adopt the values of the decay
constants: $f_{\pi}=0.130$ GeV, $f_{\rho}=0.205$ GeV,
$f_{K}=0.156$ GeV, $f_{K^*}=0.217$ GeV etc for numerical
calculations. Then the relevant decay widths for the concerned
non-leptonic decays are calculated. As the final results, we
present the decay widths by our method and the others' methods
else in Table~II. Note that the uncertainties in Table~II are
estimated as done in the previous subsection for semileptonic
decays.

For comparison precisely with the other approaches and
experimental measurements in future, we take the values $a_1=1.14$
for non-leptonic decays as done in most references, and the
experimental value of $B_c$ lifetime ${\tau}_{B_c}=0.453$ ps as
well, we calculate branching ratios of the decays and put them in
Table~III.

\begin{table}\caption{Branching ratios (in \%) of $B_c$ decays calculated for
the $B_c$ lifetime ${\tau}_{B_c}=0.453$ ps and $a_1=1.14$.}
\begin{center}
\begin{tabular}{|c|c|c|c|}
\hline\hline
Decay& Br & Decay& Br\\
\hline
$B_c^+\to{\chi}_{c0}e\nu$&$0.13\pm0.03$&$B_c^+\to{\chi}_{c0}\tau\nu$&$0.016\pm0.008$\\
$B_c^+\to{\chi}_{c1}e\nu$&$0.11\pm0.03$&$B_c^+\to{\chi}_{c1}\tau\nu$&$0.0097\pm0.0065$\\
$B_c^+\to{\chi}_{c2}e\nu$&$0.10\pm0.03$&$B_c^+\to{\chi}_{c2}\tau\nu$&$0.0082\pm0.0048$\\
$B_c^+\to h_{c}e\nu$&$0.28\pm0.08$&$B_c^+\to h_{c}\tau\nu$&$0.019\pm0.013$\\
 \hline
$B_c^+\to{\chi}_{c0}{\pi}^+$&$0.031\pm0.004$&$B_c^+\to{\chi}_{c0}{\rho}^+$&$0.076\pm0.009$\\
$B_c^+\to{\chi}_{c1}{\pi}^+$&$0.0021\pm0.0002$&$B_c^+\to{\chi}_{c1}{\rho}^+$&$0.023\pm0.002$\\
$B_c^+\to{\chi}_{c2}{\pi}^+$&$0.021\pm0.005$&$B_c^+\to{\chi}_{c2}{\rho}^+$&$0.056\pm0.011$\\
 $B_c^+\to h_c{{\pi}^+}$&$0.098\pm0.015$& $B_c^+\to h_c{{\rho}^+}$&$0.22\pm0.04$\\
\hline
$B_c^+\to{\chi}_{c0}{K}^+$&$0.0023\pm0.0003$&$B_c^+\to{\chi}_{c0}{K}^{*+}$&$0.0045\pm0.0006$\\
$B_c^+\to{\chi}_{c1}{K}^+$&$0.00016\pm0.00002$&$B_c^+\to{\chi}_{c1}{K}^{*+}$&$0.0017\pm0.0001$\\
$B_c^+\to{\chi}_{c2}{K}^+$&$0.0016\pm0.0003$&$B_c^+\to{\chi}_{c2}{K}^{*+}$&$0.0033\pm0.0006$\\
$B_c^+\to h_c{{K}^+}$&$0.0074\pm0.0011$& $B_c^+\to h_c{K}^{*+}$&$0.013\pm0.002$\\
\hline \hline
\end{tabular}
\end{center}
\end{table}

\section{Discussions and conclusions}

In Sec.~IV, the form factors (Fig.~3), energy spectrums of the
charge leptons (Fig.~4 and Fig.~5), decay widths (Table~I) for the
semileptonic decays, and the decay widths for non-leptonic decays
(Table~II) are presented. Specially in tables some comparisons
with other approaches else are also given. Thus one may read off a
lot of interesting matters already.

Since the form factors for the semi-leptonic decays, which are
directly related to overlapping integrations of the components of
the B-S wave functions of the initial and final states as shown in
Appendix~C, are comparatively difficult to be measured, so in
Fig.~3 we show the behaviors of the form factors briefly (without
errors). Whereas the energy spectrums of the charged lepton in the
decays may be measured not so difficult, as long as the event
example is great enough and the abilities of the detector are
strong enough, and to see the differences between the spectrums of
electron and $\tau$ lepton clearly in Fig.~4 we plot the curves
with center values without theoretical uncertainties. Moreover to
see the differences between this work and the ones \cite{qwerr},
in Fig.~5 we plot the spectrums of electron obtained by this work
vs the ones \cite{qwerr} obtained by previous approach and for
both of them only center value without theoretical uncertainties
are taken. Since the spectrums of muon ($\mu$) is very similar to
that of electron in exclusive semi-leptonic decays, thus we do not
present the spectrums of muon at all. From Fig.~4 we can see the
difference in the energy spectrums among the $B_c$ decays to
different $P$-wave charmonia clearly, although the results of
electron is greater than the one of $\tau$ lepton. From Fig.~5 we
can see that the difference in the energy spectrums of electron
due to different approaches: the difference caused by newly
improved approach and by the previous approach can be quite
sizable and can be tested experimentally in future. For the widths
of the decays, from Table~I and Table~II, both the semi-leptonic
decays and the non-leptonic decays, one may see that in general
the results of this work fall into the region of the predictions
by various models, but the distribution of the predictions is
quite wide, so future experimental data will be critical and may
conclude which one of the predictions is more reliable.

Considering the fact that the substantial tests of the $B_c$-meson
decays have not been started yet, although the meson $B_c$ has been
observed at Tevatron for years and LHC is running now, according to
the estimates of the production at LHC, one may believe reasonably
that the tests of the predictions on the $B_c$ decays will be
started with LHC more measurements available. From theoretical point
of view, we think that the newly improved approach works better than
the previous one, this trust need to be tested by experiments. We
would also like to note here that according to the estimates
\cite{chang-chen,z-factory,wxg1,wxg} of the production at an
$e^-e^+$ collider running at CM energy $\sqrt{S}\simeq m_Z$ ($m_Z$
is Z-boson mass) with very high luminosity
($L=10^{34\sim36}$cm$^{-2}$s$^{-1}$) i.e. a ``Super-$Z$-Factory''
and considering the advantages, may be more suitable to test the
approaches by measuring the decays precisely than that to do them at
hadronic collider such as Tevatron or LHC, because at such a
Super-$Z$-Factory numerous $B_c$ mesons may be produced and the
energy-momentum of the produced $B_c$ meson, as the $e^-e^+$ one of
the collider, is precisely known in an $e^+-e^-$ collider
environment.

\section*{\Large Acknowledgments}
This work was supported in part by Natural Science Foundation of
China (NSFC) under Grant No.10875032, No.10805082, No.10875155,
No.10847001. This research was also supported in part by the Project
of Knowledge Innovation Program (PKIP) of Chinese Academy of
Sciences, Grant No. KJCX2.YW.W10.

\appendix{

\section{The functions $\alpha$, $\beta_{++}$, $\beta_{+-}$,
$\beta_{-+}$, $\beta_{--}$, $\gamma$ }

Here according to the $P$-wave charmonium appearing in the final
state we present the useful functions $\alpha$, $\beta_{++}$,
$\beta_{+-}$, $\beta_{-+}$, $\beta_{--}$, $\gamma$ how precisely
to relate to the form factors in turn.

\noindent {\bf a). When $B_c$ decays to $\chi_{c0}$}:

Since the matrix elements of weak currents are described in terms of
two form factors ($s_+, s_-$):
\begin{eqnarray}\label{f-factor1}
&\langle{\chi}_{c0}(P_f)|V^{\mu}|B_c(P)\rangle =0\,,\nonumber\\
& \langle{\chi}_{c0}(P_f)|A^{\mu}|B_c(P)\rangle
=s_+(P+P_f)^{\mu}+s_-(P-P_f)^{\mu}\,,\nonumber
\end{eqnarray}
then the functions are read as
\begin{eqnarray}
{\beta}_{++}=s^2_+\,,\;\; {\beta}_{+-}={\beta}_{-+}=s_+s_-,\;\;
{\beta}_{--}=s^2_-\,.
\end{eqnarray}

\noindent {\bf b). When $B_c$ decays to $\chi_{c1}$}:

Since the matrix elements of weak currents can be described in terms
of four form factors ($f, u_1, u_2, g$):
\begin{eqnarray}\label{f-factor2}
&\displaystyle \langle {\chi}_{c1}(P_f)|V^{\mu}|B_c(P)\rangle
=f(M+M_f){\varepsilon}^{\mu}+[u_1P^{\mu}+u_2P_f^{\mu}]\frac{{\varepsilon}\cdot
P}{M}\,,\nonumber\\
&\displaystyle \langle {\chi}_{c1}(P_f)|A^{\mu}|B_c(P)\rangle
=\frac{2g}{M+M_f}i{\epsilon}^{\mu\nu\rho\sigma}{\varepsilon}_{\nu}P_{\rho}{P_f}_{\sigma}\,,\nonumber
\end{eqnarray}
then the functions are read as
\begin{eqnarray}
&\alpha=f_1^2+4M^2g_1^2\vec{p}_f^2\,,\nonumber\\
&\displaystyle{\beta}_{++}=\frac{f_1^2}{4M_f^2}-M^2g_1^2y+\frac{1}{2}
\left[\frac{M^2}{M_f^2}(1-y)-1\right]f_1u_+
 +\frac{M^2\vec{p}_f^2}{M_f^2}u^2_+\,,\nonumber\\
&\displaystyle\beta_{+-}=\beta_{-+}=g_1^2(M^2-M^2_f)-\frac{f_1^2}{4M^{2}_f}
-\frac{1}{2}f_1(u_{+}+u_{-})-\frac{1}{2}\frac{ME_f}{M^2_f}f_1(u_{+}-u_{-})
+u_{+}u_{-}\frac{M^2{\vec{p}_f }^2}{M^2_f}\,,\nonumber\\
&\displaystyle\beta_{--}=-g_1^2(M^2+2ME_f+M^2_f)+\frac{f_1^2}{4M^{2}_f}
-\left(\frac{ME_f}{M^2_f}+1\right)f_1u_{-}
+u^2_{-}\frac{M^2{\vec{p_f}^2}}{M^2_f}\,,\nonumber\\
&\gamma=-2f_1g_1\,
\end{eqnarray}
when setting $f_1=f(M+M_f), u_+=\frac{(u_1+u_2)}{2M},
u_-=\frac{(u_1-u_2)}{2M},g_1=\frac{g}{M+M_f}$.

\noindent {\bf c). When $B_c$ decays to $h_c$}:

Since the matrix elements of weak currents can be described in terms
of four invariant form factors ($V_0, V_1, V_2, V_3$):
\begin{eqnarray}\label{f-factor3}
&\displaystyle \langle h_c(P_f)|V^{\mu}|B_c(P)\rangle
=V_0(M+M_f){\varepsilon}^{\mu}+[V_1P^{\mu}+V_2P_f^{\mu}]\frac{{\varepsilon}\cdot
P}{M}\,,\nonumber\\
&\displaystyle \langle h_c(P_f)|A^{\mu}|B_c(P)\rangle
=\frac{2V_3}{M+M_f}i{\epsilon}^{\mu\nu\rho\sigma}{\varepsilon}_{\nu}P_{\rho}{P_f}_{\sigma}\,,\nonumber
\end{eqnarray}
then the functions are read as
\begin{eqnarray}
&\displaystyle\alpha=f_1^2+4M^2g_1^2\vec{p}_f^2\,,\nonumber\\
&\displaystyle{\beta}_{++}=\frac{f_1^2}{4M_f^2}-M^2g_1^2y+
\frac{1}{2}\left[\frac{M^2}{M_f^2}(1-y)-1\right]f_1a_+
+\frac{M^2\vec{p}_f^2}{M_f^2}a^2_+\,,\nonumber\\
&\displaystyle\beta_{+-}=\beta_{-+}=g_1^2(M^2-M^2_f)-\frac{f_1^2}{4M^{2}_f}
-\frac{1}{2}f_1(a_{+}+a_{-})-\frac{1}{2}\frac{ME_f}{M^2_f}f_1(a_{+}-a_{-})
+a_{+}a_{-}\frac{M^2{\vec{p}_f }^2}{M^2_f}\,,\nonumber\\
&\displaystyle\beta_{--}=-g_1^2(M^2+2ME_f+M^2_f)+\frac{f_1^2}{4M^{2}_f}
-\left(\frac{ME_f}{M^2_f}+1\right)f_1a_{-}+a^2_{-}\frac{M^2{\vec{p_f}^2}}{M^2_f}\,,\nonumber\\
&\gamma=-2f_1g_1\,,
\end{eqnarray}
when setting $f_1=V_0(M+M_f), a_+=\frac{(V_1+V_2)}{2M},
 a_-=\frac{(V_1-V_2)}{2M},g_1=\frac{V_3}{M+M_f}$.

\noindent {\bf d). When $B_c$ decays to $\chi_{c2}$}:

Since the matrix elements of weak currents can be described in terms
of four form factors ($k, c_1, c_2, h$):
\begin{eqnarray}\label{f-factor4}
&\displaystyle\langle{\chi}_{c2}(P_f)|A^{\mu}|B_c(P)\rangle
=k(M+M_f){\varepsilon}^{\alpha\mu}\frac{P_{\alpha}}{M}
+{\varepsilon}_{\alpha\beta}\frac{P^{\alpha}P^{\beta}}{M^2}(c_1P^{\mu}+c_2P_f^{\mu})\,,\nonumber\\
&\displaystyle\langle{\chi}_{c2}(P_f)|V^{\mu}|B_c(P)\rangle=
\frac{2h}{M+M_f}i{\varepsilon}_{\alpha\beta}\frac{P^{\alpha}}{M}
{\epsilon}^{\mu\beta\rho\sigma}P_{\rho}{P_f}_{\sigma}\,,\nonumber
\end{eqnarray}
where ${\varepsilon}_{\alpha\beta} ({\varepsilon}^{\alpha\mu})$ is
the polarization tensor of tensor meson, then the functions are read
as
\begin{eqnarray}
&\displaystyle\alpha=\frac{c}{2}(k_1^2+4Mh_1^2\vec{p}_f^{\;2})\,,\nonumber\\
&\displaystyle{\beta}_{++}=\frac{ck_1^2}{8M_f^2}-\frac{ch_1^2}{2}M^2y+\frac{2}{3}c^2c_+^2\nonumber\\
&\displaystyle+\frac{4}{3}ck_1c_+\left(\frac{M^2(1-y)+M_f^2}{4M_f^2}-\frac{1}{2}\right)
+\frac{k_1^2}{6}\left(\frac{M^2(1-y)+M_f^2}{4M_f^2}-\frac{1}{2}\right)^2\,,\nonumber\\
&\displaystyle{\beta}_{+-}={\beta}_{-+}=-\frac{ck_1^2}{8M_f^2}+\frac{ch_1^2}{2}(M^2-M_f^2)
+\frac{k_1^2}{6}\left[\left(\frac{M^2(1-y)+M_f^2}{4M_f^2}\right)^2-\frac{1}{4}\right]
+\frac{2}{3}c^2c_+c_-\nonumber\\
&\displaystyle-\frac{2}{3}ck_1c_+\left(\frac{M^2(1-y)+M_f^2}{4M_f^2}+\frac{1}{2}\right)
+\frac{2}{3}ck_1c_-\left(\frac{M^2(1-y)+M_f^2}{4M_f^2}-\frac{1}{2}\right)\,,\nonumber\\
&\displaystyle{\beta}_{--}=\frac{ck_1^2}{8M_f^2}-\frac{ch_1^2}{2}(2(M^2+M_f^2)-M^2y)
+\frac{2}{3}c^2c_-^2\nonumber\\
&\displaystyle+\frac{4}{3}ck_1c_-\left(-\frac{M^2(1-y)+M_f^2}{4M_f^2}-\frac{1}{2}\right)
+\frac{k_1^2}{6}\left(\frac{M^2(1-y)+M_f^2}{4M_f^2}+\frac{1}{2}\right)^2\,,\nonumber\\
&\gamma=-ch_1k_1\,,
\end{eqnarray}
when setting $c=\frac{M^2{\vec{p}_f}^{\;2}}{M_f^2},\,
k_1=k(1+\frac{M_f}{M}),\,c_+=\frac{c_1+c_2}{2M^2},\,
c_-=\frac{c_1-c_2}{2M^2},\,h_1=\frac{h}{M(M+M_f)}$.

\section{The B-S equation under `complete instantaneous approximation'}

In this appendix we outline the `complete instantaneous
approximation' onto the Bethe-Salpeter equation when it has an
instantaneous kernel, which describes a double heavy meson quite
well.

The Bethe-Salpeter equation \cite{BS} is read as
\begin{equation}\label{eq10}
(\not\!p_{1}-m_{1})\chi_{p}(q)(\not\!p_{2}+m_{2})=i\int\frac{d^{4}k}{(2\pi)^{4}}V(P,k,q)\chi_{p}(k),
\end{equation}
where $\chi_{p}(q)$ is B-S wave function of the relevant bound
state, $P$ is the four momentum of the meson state and $p_{1}$,
$p_{2}$, $m_{1}$, $m_{2}$ are the momenta and constituent masses of
the quark and anti-quark respectively. From the definition, they
relate to the total momentum $P$ and relative momentum $q$ as
follows:
$$p_1 = \alpha_1P +
q,\  \alpha_1 \equiv \frac{m_1}{ m_1 + m_2} ,$$
$$ p_2 = \alpha_2P-q,\  \alpha_2\equiv\frac{ m_2}{ m_1 + m_2}.$$
The interaction kernel $V(P,k,q)$ for a double heavy system, being
instantaneous approximately, can be treated as a potential after
doing instantaneous approximation, i.e. the kernel take the simple
form (in the rest frame) \cite{Salp}
$$V(P,k,q)\Rightarrow V(|\vec{k}-\vec{q}|)\,.$$

For various usages, we divide the relative momentum $q$ into two
parts,
$$q^\mu=q^\mu_\parallel+q^\mu_\perp,\ \ q^\mu_\parallel\equiv
\frac{ P\cdot q}{M^2}P^\mu,\ \ q^\mu_\perp\equiv
q^\mu-q^\mu_\parallel\,,$$ where $M$ is the mass of the meson, and
we may have two Lorentz invariant variables:
$$q_P\equiv \frac{P\cdot q}{M},\ \ q_T\equiv \sqrt{-q_\perp^2}\,.$$

For the convenience below, let us introduce the definitions:
\begin{equation}\label{BSE3} \varphi_{p}(q_\perp^{\mu})\equiv i\int
\frac{\mathrm{d}q_{p}}{2\pi}\chi
_{p}(q_\parallel^{\mu},q_\perp^{\mu}),\ \ \eta(q_\perp^{\mu})\equiv
\int
\frac{\mathrm{d}k_\perp^{3}}{(2\pi)^{3}}V(k_\perp,q_\perp)\varphi_{p}(k_\bot^{\mu})\,,
\end{equation}
then the B-S equation can be rewritten as
\begin{equation}\label{BSE2}
\chi(q_{\|},q_{\bot})=S_1(p_1)\eta(q_{\bot})S_2(p_2)\,.
\end{equation}
Owing to Eqs.~(\ref{BSE3},~\ref{BSE2}), it is reasonable and for
convenience we may call $\eta(q_{\bot})$ as `instantaneous B-S
vertex'. The propagator of quark or anti-quark may be decomposed:
$$S_i(p_i)=\frac{\Lambda_{i}^{+}(q_\perp)}{J(i)q_P+{\alpha}_iM-w_i+i\epsilon}
+\frac{\Lambda_{i}^{-}(q_\perp)}{J(i)q_P+{\alpha}_iM-w_i+i\epsilon}\,,$$
where $i$=1,~2 for quark and anti-quark respectively, and
$J(i)=(-1)^{i+1}$, $\omega_{1}=\sqrt{m_{1}^{2}+q_T^{2}}$,
$\omega_{2}=\sqrt{m_{2}^{2}+q_T^{2}}$, and $\Lambda_{1}^{\pm},\
\Lambda_{2}^{\pm}$ are the generalized energy projection
operators,
\begin{eqnarray}\label{projector1}
&\displaystyle\Lambda_{1}^{\pm}(q_\perp)\equiv
\frac{1}{2\omega_{1}}[\frac{\not\!{P}}{M}\omega_{1}\pm(m_{1}+\not\!q_\perp)],\
\ \ \  \Lambda_{2}^{\pm}(q_\perp)\equiv
\frac{1}{2\omega_{2}}[\frac{\not\!{P}}{M}\omega_{2}\mp(m_{2}+\not\!q_\perp)]\,,
\end{eqnarray}
and have the properties:
\begin{eqnarray}\label{properties}
&\displaystyle\Lambda^{+}_{iP}(q^\mu_{P_\perp})+
\Lambda^{-}_{iP}(q^\mu_{P_\perp})=
\frac{\not\!{P}}{M}\,,\;\;\;\;\;\;\;\;\;\;\;
\Lambda^{\pm}_{iP}(q^\mu_{P_\perp})\frac{\not\!{P}}{M}
\Lambda^{\mp}_{iP}(q^\mu_{P_\perp})=0 \,,\nonumber \\
&\displaystyle\Lambda^{\pm}_{iP}(q^\mu_{P_\perp})\frac{\not\!{P}}{M}
\Lambda^{\pm}_{iP}(q^\mu_{P_\perp})
=\Lambda^{\pm}_{iP}(q^\mu_{P_\perp})\,,
\end{eqnarray}

The instantaneous approximation to the B-S equation is to do
contour integration over $q_P$ on both sides of Eq.~(\ref{BSE2}),
and obtains:
\begin{equation}\label{Salp0}
\varphi_{p}(q_\perp)=\frac{\Lambda_{1}^{+}(q_\perp)\eta(q_\perp)
\Lambda_{2}^{+}(q_\perp)}{M-\omega_{1}-\omega_{2}}-\frac{\Lambda_{1}^{-}(q_\perp)\eta(q_\perp)
\Lambda_{2}^{-}(q_\perp)}{M+\omega_{1}+\omega_{2}}\,,
\end{equation}
If we introduce the notations:
\begin{equation}\label{projector}
\varphi_{p}^{\pm\pm}(q_\perp)\equiv\Lambda_{1}^{\pm}(q_\perp)
\frac{\not\!{P}}{M}\varphi_{p}(q_\perp)\frac{\not\!{P}}{M}\Lambda_{2}^{\pm}(q_\perp)\,,
\end{equation}
we have
\begin{equation}
\varphi_{p}(q_\perp)=\varphi_{p}^{++}(q_\perp)+\varphi_{p}^{+-}(q_\perp)
+\varphi_{p}^{-+}(q_\perp)+\varphi_{p}^{--}(q_\perp)\,,
\end{equation}
With the properties Eq.~(\ref{properties}) and notations
Eq.~(\ref{projector}), the full Salpeter equation
Eq.~(\ref{Salp0}) can be written as
\begin{eqnarray}\label{Salp10}
(M-\omega_{1}-\omega_{2})\varphi_{p}^{++}(q_\perp)=\Lambda_{1}^{+}(q_\perp)\eta(q_\perp)
\Lambda_{2}^{+}(q_\perp)\,,
\end{eqnarray}
\begin{eqnarray}\label{Salp11}
(M+\omega_{1}+\omega_{2})\varphi_{p}^{--}(q_\perp)=-\Lambda_{1}^{-}(q_\perp)\eta(q_\perp)
\Lambda_{2}^{-}(q_\perp)\,,
\end{eqnarray}
\begin{eqnarray}\label{Salp12}
&&\varphi_{p}^{+-}(q_\perp)=\varphi_{p}^{-+}(q_\perp)=0\,.
\end{eqnarray}
The normalization condition for the B-S equations now is read as:
\begin{equation}\label{norm}
\int\frac{q^2_Tdq_T}{2{\pi}^2}Tr[\bar{\varphi}^{++}\frac{\not\!{P}}{M}
{\varphi}^{++}\frac{\not\!{P}}{M}
-\bar{\varphi}^{--}\frac{\not\!{P}}{M}
{\varphi}^{--}\frac{\not\!{P}}{M}]=2P_0\,.
\end{equation}
The couple equations Eq.~(\ref{Salp10}), Eq.~(\ref{Salp11}) and
Eq.~(\ref{Salp12}) with the normalization condition
Eq.~(\ref{norm}) are the final B-S (Salpeter) equation under
`complete instantaneous approximation' vs the previous one i.e.
Salpeter equation \cite{Salp} where only Eq.~(\ref{Salp10}) is
considered.

In addition, note that in the model used here for the double heavy
quark-antiquark systems, the QCD-inspired interaction kernel $V$,
being instantaneous approximately and dictating the Cornell
potential which is composed by a linear scalar interaction plus a
vector interaction, is read as:
\begin{eqnarray}
&\displaystyle V(\vec{q})=V_s(\vec{q})+V_v(\vec{q})\gamma^0\otimes\gamma^0,\nonumber\\
&\displaystyle
V_s(\vec{q})=-(\frac{\lambda}{\alpha}+V_0)\delta^3(\vec{q})
+\frac{\lambda}{\pi^2}\frac{1}{(\vec{q}^2+\alpha^2)^2},\nonumber\\
&\displaystyle
V_v(\vec{q})=-\frac{2}{3\pi^2}\frac{\alpha_s(\vec{q})}{(\vec{q}^2+\alpha^2)},
\end{eqnarray}
where the QCD running coupling constant
$\alpha_s(\vec{q})=\frac{12\pi}{33-2N_f}\frac{1}{\mathrm{log}(a+\vec{q}^2/\Lambda_{QCD}^2)}$;
the constants $\lambda,\ \alpha,\ a,\ V_0$ and $\Lambda_{QCD}$ are
the parameters characterizing the potential.

\section{The reduced wave functions ${\varphi}^{++}(\vec{q})$
and the form factors}

In the appendix we present the reduced wave functions
${\varphi}^{++}(\vec{q})$ (and ${\psi}^{+-}(\vec{q})$) which
directly relate to the solutions by newly solving the obtained
coupled equations Eq.~(\ref{Salp10}), Eq.~(\ref{Salp11}) and
Eq.~(\ref{Salp12}) under a new approach. The key point of the new
approach is to solve the B-S equation according to the quantum
numbers of the concerned bound states respectively
\cite{spectr,glwang,wanggl}, i.e. to solve the equation under the
new approach we need to give the most general formulation for the
wave function first. Therefore for the present usage, in this
appendix, we precisely quote the solutions for the low-laying
bound states $B_c$ meson with quantum numbers $J^P=0^-$,
$\chi_{c0}$ with quantum numbers $J^{PC}=0^{++}$, $\chi_{c1}$ with
quantum numbers $J^{PC}=1^{++}$, $\chi_{c2}$ with quantum numbers
$J^{PC}=2^{++}$ and $h_{c}$ with quantum numbers $J^{PC}=1^{+-}$
from \cite{spectr,glwang,wanggl}, and then we write down the
reduced wave functions ${\varphi}^{++}(\vec{q})$ and the form
factors accordingly.

When the weak-current matrix elements are computed precisely, as an
intermediate step, the form factors can be represented as
overlapping integrations of the components appearing in the B-S
solutions, thus in this appendix we also give the formulas of the
form factors in terms of the `overlapping integrations'.

\noindent {\bf a). For $B_c$ meson with quantum numbers
$J^{P}=0^{-}$}

The B-S wave function (solution of Eq.~(\ref{Salp10}),
Eq.~(\ref{Salp11}) and Eq.~(\ref{Salp12}) of $B_c$ meson with
$J^P=0^-$ is read as:
\begin{equation}\label{bc1}
{\varphi}_{_{B_c}}(\vec{q})=M\Bigg[\frac{\not\!{P}}{M}{f}_{1}(\vec{q})\Big\{1-
\frac{\not\!{q_{\bot}}(w_1+w_2)}{m_2w_1+m_1w_2}\Big\}+{f}_{2}(\vec{q})\Big\{1
+\frac{\not\!q_{\bot}(w_2-w_1)}{m_1w_2+m_2w_1}\Big\}\Bigg]{\gamma}_5,
\end{equation}
where M, $P$ are the mass and the total momentum of the meson $B_c$,
$q_{\bot}=(0,\vec{q})$, $\vec{q}$ is the relative momentum of quark
and anti-quark in the meson, so $q_{\bot}^2=-{\vec{q}}^{~2}$.

Then we can rewrite the reduced wave function:
\begin{eqnarray}
{\varphi}^{++}_{_{B_c}}(\vec{q})=b_1
\left[b_2+\frac{\not\!{P}}{M}+b_3\not\!{q_{\bot}}
+b_4\frac{\not\!{q_{\bot}}\not\!{P}}{M}\right]{\gamma}_5,
\end{eqnarray}
where
$$b_1=\frac{M}{2}\left({f}_{1}(\vec{q})
+{f}_{2}(\vec{q})\frac{m_1+m_2}{w_1+w_2}\right),\;\;\;\;
b_2=\frac{w_1+w_2}{m_1+m_2},$$
$$b_3=-\frac{(m_1-m_2)}{m_1w_2+m_2w_1},\;\;\;\;
b_4=\frac{(w_1+w_2)}{(m_1w_2+m_2w_1)}.$$

In Appendix.~B in Eq.~(\ref{BSE3}), we have
\begin{eqnarray}
&\displaystyle \eta(q_{\bot})=\int
d^3kV(\vec{k})M\Bigg[\frac{\not\!{P}}{M}{f}_{1}(\vec{k})\Big\{1-
\frac{\not\!{k_{\bot}}(w_{11}+w_{21})}{m_2w_{11}+m_1w_{21}}\Big\}\nonumber\\
&\displaystyle +{f}_{2}(\vec{k})\Big\{1
+\frac{\not\!k_{\bot}(w_{21}-w_{11})}{m_1w_{21}+m_2w_{11}}\Big\}\Bigg]{\gamma}_5,
\end{eqnarray}
where $w_{11}=\sqrt{m_1^2-k_{\bot}^2}$,
$w_{21}=\sqrt{m_2^2-k_{\bot}^2}$,
$V(\vec{k})=V_s(\vec{k})+V_v(\vec{k})\gamma^0\otimes\gamma^0$.

 According to Eq.~(\ref{bc1}),
\begin{eqnarray}
&\displaystyle \eta(q_{\bot})=\int
d^3k(V_s(\vec{k})+V_v(\vec{k})\gamma^0\otimes\gamma^0)\nonumber\\
&\displaystyle  M\Bigg[\frac{\not\!{P}}{M}{f}_{1}(\vec{k})\Big\{1-
\frac{\not\!{k_{\bot}}(w_{11}+w_{21})}{m_2w_{11}+m_1w_{21}}\Big\}+{f}_{2}(\vec{k})\Big\{1
+\frac{\not\!k_{\bot}(w_{21}-w_{11})}{m_1w_{21}+m_2w_{11}}\Big\}\Bigg]{\gamma}_5\nonumber\\
&\displaystyle
=M\left[g_1\frac{\not\!{P}}{M}+g_2+g_3\frac{\not\!{q_{\bot}}}{M}+g_4\frac{\not\!{P\not\!{q_{\bot}}}}{M^2}\right]\gamma_5.
\end{eqnarray}

$$g_1=\int d^3k[V_s-V_v]f_1(\vec{k}),\ \ g_2=\int d^3k[V_s-V_v]f_2(\vec{k}),$$

$$g_3=\int d^3k[V_s+V_v]\frac{\vec{k}\cdot\vec{q}}{|\vec{q}|^2}f_2(\vec{k})
\frac{(w_{21}-w_{11})}{m_1w_{21}+m_2w_{11}},\ \ g_4=\int
d^3k[V_s+V_v]f_1(\vec{k})\frac{(w_{11}+w_{21})}{m_1w_{21}+m_2w_{11}}.$$

So we can also write down the wave function of
${\psi}^{+-}(q_\perp)$,
\begin{eqnarray}
{\psi}^{+-}(q_\perp)=
\frac{\Lambda^{+}_{1}(q_{P\perp})\eta(q_\perp)\Lambda^{-}_{2}(q_{P\perp})
}{M+\omega_{2}+\omega'_{2}-E_f}
=\left[n_1\frac{\not\!{P}}{M}+n_2+n_3\not\!{q_{\bot}}
+n_4\frac{\not\!{q_{\bot}}\not\!{P}}{M}\right]{\gamma}_5.
\end{eqnarray}
set $tt=\frac{1}{4w_1w_2(M+\omega_{2}+\omega'_{2}-E_f)}$, where
the symbol $'$, denotes of the final state, and
$$n_1=tt[g_1M(-q^2+m_1m_2-w_1w_2)+g_2M(m_2w_1-m_1w_2)+g_3(w_1+w_2)q^2+g_4(m_1+m_2)q^2],$$
$$n_2=tt[g_1M(m_2w_1-m_1w_2)+g_2M(q^2+m_1m_2-w_1w_2)+g_3(m_1-m_2)q^2+g_4(w_1-w_2)q^2],$$
$$n_3=tt[-g_1M(w_1+w_2)-g_2M(m_1-m_2)+g_3(q^2+m_1m_2+w_1w_2)+g_4(m_2w_1+m_1w_2)],$$
$$n_4=tt[g_1M(m_1+m_2)+g_2M(w_1-w_2)-g_3(m_2w_1+m_1w_2)-g_4(-q^2+m_1m_2+w_1w_2)].$$

\noindent{\bf b). For the charmonium $\chi_{c0}~(J^{PC}=0^{++})$
and the form factors $s_+$ and $s_-$}

The B-S wave function (solution of Eq.~(\ref{Salp10}),
Eq.~(\ref{Salp11}) and Eq.~(\ref{Salp12}) under new method to
solve the coupled equations) of $\chi_{c0}$ is read as:
\begin{equation}
{\varphi}_{\chi_{c0}}(\vec{q'})={f}'_{1}(\vec{q'})\not\!{q'_{\bot}}+{f}'_{2}(\vec{q'})\frac{\not\!{P_f}\not\!{q'_{\bot}}}{M_f}
+{f}'_{3}(\vec{q'})M_f+{f}'_{4}(\vec{q'})\not\!{P_f}\,,
\end{equation}
with constraints on the components of wave function, for the
charmonium, $m'_1=m'_2, w'_1=w'_2$, we get:
$$f'_3(\vec{q'})=\frac{f'_1(\vec{q'})q_{\bot}^{\prime2}}{M_fm'_1},\;\;\;
f'_4(\vec{q'})=0\,,$$ where $M_f, P_f$ are the mass and the total
momentum of final meson $\chi_{c0}$, $q'_{\bot}=(0,\vec{q'})$,
$\vec{q'}$ is the relative momentum of quark and anti-quark in the
meson, so $q_{\bot}^{\prime2}=-{\vec{q'}}^{~2}$. Then the reduced
wave function ${\varphi}_{^3P_0}^{++}(\vec{q'})$ as:
\begin{equation}
{\varphi}_{\chi_{c0}}^{++}(\vec{q'})=a_1\left[\not\!{q'_{\bot}}+a_2\frac{\not\!{P_f}\not\!{q'_{\bot}}}{M_f}
+a_3+a_4\frac{\not\!{P_f}}{M_f}\right]\,,
\end{equation}
with $$a_1=\frac{1}{2}\left({f}'_{1}(\vec{q'})
+{f}'_{2}(\vec{q'})\frac{m'_1}{w'_1}\right),\;\;
a_2=\frac{w'_1}{m'_1},\ \ \
a_3=\frac{q_{\bot}^{\prime2}}{m'_1},\;\;a_4=0.$$

The wave function of $\bar{\psi}^{\prime-+}(q'_{P\perp})$ is
\begin{eqnarray}
\bar{\psi}^{\prime-+}(q'_{P\perp})=
\frac{\Lambda'^{-}_2(q'_{P\perp})
\bar{\eta}'(q'_{P\perp})\Lambda'^{+}_1(q'_{P\perp})}
{M-\omega_{2}-\omega'_{2}-E_f}=
n'_1\not\!{q'_{\bot}}+n'_2\frac{\not\!{q'_{\bot}}\not\!{P_f}}{M_f}
+n'_3+n'_4\frac{\not\!{P_f}}{M_f}.
 \end{eqnarray}
Set $tt'=\frac{1}{4w_1^{\prime2}(M-\omega_{2}-\omega'_{2}-E_f)}$,
where
$$n'_1=tt'[-2g'_1q'^2+2g'_3M_fm'_1],\ \ \ n'_2=0,$$
$$n'_3=tt'[-2g'_1m'_1q'^2+2g'_3M_fm_1^{\prime2}], \ \ \ n'_4=tt'[-2g'_1w'_1q'^2+2g'_3M_fm'_2w'_1],$$ and
$$g'_1=\int d^3k'[V_s-V_v]\frac{\vec{k'}\cdot\vec{q'}}{|\vec{q'}|^{2}}f'_1(\vec{k'}),
\ \ g'_2=\int
d^3k'[V_s-V_v]\frac{\vec{k'}\cdot\vec{q'}}{|\vec{q'}|^2}f'_2(\vec{k'}),$$

$$g'_3=\int
d^3k'[V_s+V_v]\frac{f'_1(\vec{k'})k_{\bot}^{\prime 2}}{M_fm'_1}, \ \
g'_4=0.$$
 With Eq.~(\ref{previous}), the form
factors may be presented by overlapping integrations:
\begin{eqnarray}
&&s_+=\frac{1}{2}\int\frac{d^3q}{(2\pi)^3}\frac{4a_1b_1}{MM_f}
\left[a_3b_2M_f+{\alpha}_{11}E_f(a_2b_2E_f+M_f+a_2b_4\vec{q}\cdot\vec{P_f})\right.\nonumber\\
&&+b_3(M_fq^2+{\alpha}_{11}M_f\vec{q}\cdot\vec{P_f})
+M(a_2b_4q^2-{\alpha}_{11}a_2b_2E_f-{\alpha}_{11}M_f)\nonumber\\
&&\left.+M\frac{q\cos\theta}{|\vec{P_f}|}
(1-\frac{E_f}{M})(a_2b_2E_f-a_3b_4M_f+M_f+a_2b_4\vec{q}\cdot\vec{P_f})\right],
\end{eqnarray}
\begin{eqnarray}
&&s_-=\frac{1}{2}\int\frac{d^3q}{(2\pi)^3}\frac{4a_1b_1}{MM_f}
\left[a_3b_2M_f+{\alpha}_{11}E_f(a_2b_2E_f+M_f+a_2b_4\vec{q}\cdot\vec{P_f})\right.\nonumber\\
&&+b_3(M_fq^2+{\alpha}_{11}M_f\vec{q}\cdot\vec{P_f})
-M(a_2b_4q^2-{\alpha}_{11}a_2b_2E_f-{\alpha}_{11}M_f)\nonumber\\
&&\left.-M\frac{q\cos\theta}{|\vec{P_f}|}
(1+\frac{E_f}{M})(a_2b_2E_f-a_3b_4M_f+M_f+a_2b_4\vec{q}\cdot\vec{P_f})\right],
 \end{eqnarray}
where ${\alpha}_{11}=\alpha'_1=\frac{m'_1}{m'_1+m'_2}.$

\noindent{\bf c). For the charmonium $\chi_{c1}~ (J^{PC}=1^{++})$
and form factors the $f$, $u_1$, $u_2$, $g$}

The B-S wave function (solution of Eq.~(\ref{Salp10}),
Eq.~(\ref{Salp11}) and Eq.~(\ref{Salp12}) under new method to
solve the coupled equations) of $\chi_{c1}$ is read as:
\begin{equation}
{\varphi}_{\chi_{c1}}(\vec{q'})=i{\epsilon}_{\mu\nu\alpha\beta}P_f^{\nu}q'^{\alpha}_{\bot}{\varepsilon}^{\beta}
[f'_1(\vec{q'})M_f{\gamma}^{\mu}+f'_2(\vec{q'})\not\!{P_f}{\gamma}^{\mu}+f'_3(\vec{q'})\not\!{q'_{\bot}}{\gamma}^{\mu}
+if'_4(\vec{q'}){\epsilon}^{\mu\rho\sigma\delta}P_{f\sigma}q'_{\bot\rho}{\gamma}_{\delta}{\gamma}_5/M_f]/M_f^2\,,
\end{equation}
where $\varepsilon$ is the polarization vector of axial vector meson
and with the constraint on the components:
$$f'_3(\vec{q'})=0,\;\;\;
 f'_4(\vec{q'})=\frac{f'_2(\vec{q'})M_f}{m'_1},$$
Then the reduced wave function ${\varphi}_{^3P_1}^{++}(\vec{q'})$
as:
\begin{equation}
{\varphi}_{\chi_{c1}}^{++}(\vec{q'})=i{\epsilon}_{\mu\nu\alpha\beta}P_f^{\nu}q'^{\alpha}_{\bot}{\varepsilon}^{\beta}
a_1[M_f{\gamma}^{\mu}+a_2{\gamma}^{\mu}\not\!{P_f}+a_3{\gamma}^{\mu}\not\!{q'_{\bot}}
+a_4{\gamma}^{\mu}\not\!{P_f}\not\!{q'_{\bot}}]/M_f^2\,,
\end{equation}
with
$$a_1=\frac{1}{2}\left(f'_1(\vec{q'})+f'_2(\vec{q'})\frac{w'_1}{m'_1}\right),\;\;\;\;
 a_2=-\frac{m'_1}{w'_1},\ \ \ a_3=0,\;\;\;\;
 a_4=-\frac{1}{w'_1}.$$

With Eq.~(\ref{previous}), the form factors may be presented by
overlapping integrations:

\begin{eqnarray}
&\displaystyle
f=\int\frac{d^3q}{(2\pi)^3}\frac{4a_1b_1}{M_f^2(M+M_f)}
\left[\left(a_4M_f^2q^2-a_2b_4M_f^2q^2
-a_2b_2E_f\vec{q}\cdot\vec{P_f}\right.\right.\nonumber\\
&\displaystyle
+(a_4-a_2b_4)(\vec{q}\cdot\vec{P_f})^2+{\alpha}_{11}^2a_4E_f^2\vec{P_f}^2+a_2b_2{\alpha}_{11}(E_f^3-E_fM_f^2)
-2{\alpha}_{11}a_4E_f^2\vec{q}\cdot\vec{P_f}
\nonumber\\
&\displaystyle
+{\alpha}_{11}a_2b_4E_f^2\vec{q}\cdot\vec{P_f}+b_3E_fM_f(q^2-{\alpha}_{11}\vec{q}\cdot\vec{P_f})
\left.\right)
+\frac{q^2}{2}(\cos^2\theta-1)\left(M_f^2(a_4-a_2b_4)\right.
\nonumber\\
&\displaystyle \left.\left.+b_3E_f(M_f+a_4\vec{q}\cdot\vec{P_f}
-{\alpha}_{11}a_4\vec{P_f}^2)\right)\right],
\end{eqnarray}
\begin{eqnarray}
&\displaystyle
u_1=\int\frac{d^3q}{(2\pi)^3}\frac{4a_1b_1M}{M_f^2}\left[\frac{{\alpha}_{11}E_f}{M^2}\left(a_2b_2M_f^2
+M_fb_3\vec{q}\cdot\vec{P_f}\right.\right.\nonumber\\
&\displaystyle \left.+a_4({\alpha}_{11}E_fM_f^2+b_3(M_f^2q^2
+(\vec{q}\cdot\vec{P_f})^2-{\alpha}_{11}\vec{q}\cdot\vec{P_f}E_f^2))\right)-{\alpha}_{11}E_f^2\frac{q\cos\theta}{M^2|\vec{P_f}|}
\left((a_4-a_2b_4)M_f^2\right.\nonumber\\
&\displaystyle \left.
+b_3E_f(-{\alpha}_{11}a_4E_f^2+M_f+a_4\vec{q}\cdot\vec{P_f}+{\alpha}_{11}a_4M_f^2)\right)\nonumber\\
&\displaystyle
-\frac{E_f}{M^2}\frac{q\cos\theta}{|\vec{P_f}|}\left(a_2b_2M_f^2
+M_fb_3\vec{q}\cdot\vec{P_f}\right.
\left.+a_4({\alpha}_{11}E_fM_f^2+b_3(M_f^2q^2
+(\vec{q}\cdot\vec{P_f})^2-{\alpha}_{11}\vec{q}\cdot\vec{P_f}E_f^2))\right)\nonumber\\
&\displaystyle
+\frac{q^2}{2M^2|\vec{P_f}|^2}(-M_f^2+(2E_f^2+M_f^2)\cos^2\theta)\left(M_f^2(a_4-a_2b_4)\right.
\nonumber\\
&\displaystyle \left.\left.+b_3E_f(M_f+a_4\vec{q}\cdot\vec{P_f}
-{\alpha}_{11}a_4\vec{P_f}^2)\right)\right],
\end{eqnarray}
\begin{eqnarray}
&\displaystyle
u_2=\int\frac{d^3q}{(2\pi)^3}\frac{4a_1b_1M}{M_f^2}\left[-\frac{1}{M}\left(b_3M_fq^2
+(a_2b_2+{\alpha}_{11}a_4E_f)({\alpha}_{11}E_f^2-\vec{q}\cdot\vec{P_f})\right)\right.
\nonumber\\
&\displaystyle
+\frac{E_fq\cos\theta}{M|\vec{P_f}|}\left({\alpha}_{11}M_fb_3E_f
+(a_2b_4-a_4)(\vec{q}\cdot\vec{P_f}-{\alpha}_{11}E_f^2)\right)\nonumber\\
&\displaystyle
+\frac{1}{M}\frac{q\cos\theta}{|\vec{P_f}|}\left(a_2b_2M_f^2
+M_f(b_3\vec{q}\cdot\vec{P_f}\right.\nonumber\\
&\displaystyle
\left.+{\alpha}_{11}b_4E_f\vec{q}\cdot\vec{P_f})+a_4({\alpha}_{11}E_fM_f^2+b_3(M_f^2q^2
+(\vec{q}\cdot\vec{P_f})^2-{\alpha}_{11}\vec{q}\cdot\vec{P_f}E_f^2))\right)\nonumber\\
&\displaystyle
-\frac{E_fq^2}{2M|\vec{P_f}|^2}(3\cos^2\theta-1)\left(M_f^2(a_4-a_2b_4)
\left.+b_3E_f(M_f+a_4\vec{q}\cdot\vec{P_f}
-{\alpha}_{11}a_4\vec{P_f}^2)\right)\right],
\end{eqnarray}
\begin{eqnarray}
&\displaystyle
g=\int\frac{d^3q}{(2\pi)^3}\frac{4a_1b_1(M+M_f)}{M_f^2}
\left[\frac{E_f}{M}(\frac{q\cos\theta}{|\vec{P_f}|}-{\alpha}_{11})\left(a_2(b_2E_f+b_4\vec{q}\cdot\vec{P_f})+M_f
\right.\right.\nonumber\\
&\displaystyle
\left.\left.+a_4b_3E_f(q^2-{\alpha}_{11}\vec{q}\cdot\vec{P_f})
\right)+\frac{q^2}{2M}(\cos^2\theta-1)
\left(b_3(M_f+a_4(\vec{q}\cdot\vec{P_f}-{\alpha}_{11}E_f^2))\right)\right].
\end{eqnarray}

\noindent{\bf d). For the charmonium $h_{c} ~(J^{PC}=1^{+-})$ and
form factors the $V_0$, $V_1$, $V_2$, $V_3$ }

The B-S wave function (solution of Eq.~(\ref{Salp10}),
Eq.~(\ref{Salp11}) and Eq.~(\ref{Salp12}) under new method to
solve the coupled equations) of $h_{c}$ is read as:
\begin{equation}
{\varphi}_{h_c}(\vec{q'})=q'_{\bot}\cdot\varepsilon\left[f'_1(\vec{q'})+f'_2(\vec{q'})\frac{\not\!P_f}{M_f}
+f'_3(\vec{q'})\not\!{q'_{\bot}}+f'_4(\vec{q'})\frac{\not\!P_f\not\!{q'_{\bot}}}{M_f^2}\right]{\gamma}_5,
\end{equation}
with the constraint on the components of the wave function,
$$f'_3(\vec{q'})=0,\;\;\;
 f'_4(\vec{q'})=-\frac{f'_2(\vec{q'})M_f}{m'_1},$$
Then we have the reduced wave function
${\varphi}_{h_c}^{++}(\vec{q'})$:
\begin{equation}
{\varphi}_{h_c}^{++}(\vec{q'})=q'_{\bot}\cdot\varepsilon
a_1\left[1+a_2\frac{\not\!P_f}{M_f}
+a_3\not\!{q'_{\bot}}+a_4\frac{\not\!{q'_{\bot}}\not\!P_f}{M_f}\right]{\gamma}_5,
\end{equation}
$$a_1=\frac{1}{2}\left(f'_1(\vec{q'})+f'_2(\vec{q'})\frac{w'_1}{m'_1})\right)\,,\;\;
a_2=\frac{m'_1}{w'_1}\,,\ \ a_3=0\,,\;\;\; a_4=\frac{1}{w'_1}\,.$$

With Eq.~(\ref{previous}), the form factors may be presented by
overlapping integrations:
\begin{eqnarray}
&&V_0=\int\frac{d^3q}{(2\pi)^3}\frac{4a_1b_1}{M_f(M+M_f)}\frac{q^2}{2}(\cos^2\theta-1)\nonumber\\
&&\left[a_4b_2E_f+a_2b_3E_f-b_4M_f+a_4b_4\vec{q}\cdot\vec{P_f}\right],
\end{eqnarray}
\begin{eqnarray}
&\displaystyle
V_1=\int\frac{d^3q}{(2\pi)^3}\frac{4a_1b_1}{M_f}\left[\frac{E_f}{M}({\alpha}_{11}-\frac{q\cos\theta}{|\vec{P_f}|})
\left({\alpha}_{11}a_4b_2E_f^2+{\alpha}_{11}a_4b_4E_f\vec{q}\cdot\vec{P_f}+b_2M_f+a_2b_3\vec{q}\cdot\vec{P_f}\right)\right.\nonumber\\
&\displaystyle
\left(\frac{q^2}{2M|\vec{P_f}|^2}(-M_f^2+(2E_f^2+M_f^2)\cos^2\theta)
-{\alpha}_{11}\frac{E_f^2q\cos\theta}{M|\vec{P_f}|}\right)\nonumber\\
&\displaystyle
\left.\left(a_4b_2E_f+a_2b_3E_f-b_4M_f+a_4b_4\vec{q}\cdot\vec{P_f}\right)\right],
\end{eqnarray}
\begin{eqnarray}
&\displaystyle
V_2=\int\frac{d^3q}{(2\pi)^3}\frac{4Ma_1b_1}{M_f}\left[\frac{E_f}{M}({\alpha}_{11}-\frac{q\cos\theta}{|\vec{P_f}|})
\left(-a_4b_4q^2+a_2-a_4b_2E_f{\alpha}_{11}\right)\right.\nonumber\\
&\displaystyle+({\alpha}_{11}\frac{E_fq\cos\theta}{M|\vec{P_f}|}+\frac{E_fq^2}{2M|\vec{P_f}|^2}(3\cos^2\theta-1))
\left.\left(a_4b_2E_f+a_2b_3E_f-b_4M_f+a_4b_4\vec{q}\cdot\vec{P_f}\right)\right],
\end{eqnarray}
\begin{eqnarray}
&\displaystyle
V_3=-\int\frac{d^3q}{(2\pi)^3}\frac{4a_1b_1(M+M_f)}{MM_f}\frac{q^2}{2}(\cos^2\theta-1)
\left[a_4(b_2+b_4E_f{\alpha}_{11})+b_3a_2\right].
\end{eqnarray}

\noindent{\bf e). For the charmonium $\chi_{c2}~ (J^{PC}=2^{++})$
and form factors the $k$, $c_1$, $c_2$, $h$ }

The B-S wave function (solution of Eq.~(\ref{Salp10}),
Eq.~(\ref{Salp11}) and Eq.~(\ref{Salp12}) under new method to
solve the coupled equations) of $\chi_{c2}$ is read as:
\begin{eqnarray}
&\displaystyle {\varphi}_{\chi_{c2}}(\vec{q'})={\varepsilon}_{\mu
\nu}q'^{\nu}_{\bot}\{q'^{\mu}_{\bot}[f'_1(\vec{q'})+\frac{\not\!P_f}{M_f}f'_2(\vec{q'})
+\frac{\not\!q'_{\bot}}{M_f}f'_3(\vec{q'})
+\frac{\not\!P_f\not\!q'_{\bot}}{M_f^2}f'_4(\vec{q'})]\nonumber\\
&\displaystyle
+{\gamma}^{\mu}[M_ff'_5(\vec{q'})+\not\!P_ff'_6(\vec{q'})+\not\!q'_{\bot}f'_7(\vec{q'})]+
\frac{i}{M_f}f'_8(\vec{q'}){\epsilon}^{\mu\alpha\beta\delta}
P_{f\alpha}q'_{\bot\beta}{\gamma}_{\delta}{\gamma}_5\}\,,
\end{eqnarray}
with the constraint on the components of the wave function:
$$f'_1(\vec{q'})=\frac{[q_{\bot}^{\prime2}f'_3(\vec{q'})+M_f^2f'_5(\vec{q'})]}{M_fm'_1}\,,\ \
f'_2(\vec{q'})=0\,,\;\;\; f'_7=0\,,\;\;\;
f'_8=\frac{f'_6(\vec{q'})M_f}{m'_1}\,,$$ where ${\varepsilon}_{\mu
\nu}$ is a tensor for $J=2$. Then we have the reduced wave function
${\varphi}_{\chi_{c2}}^{++}(\vec{q})$ as:
\begin{eqnarray}
&\displaystyle{\varphi}^{++}_{\chi_{c2}}(\vec{q'})={\varepsilon}_{\mu
\nu}q'^{\nu}_{\bot}\{q'^{\mu}_{\bot}[a_1+a_2\frac{\not\!P_f}{M_f}
+a_3\frac{{\not\!q'_{\bot}}}{M_f}\nonumber\\
&\displaystyle +a_4\frac{\not\!q'_{\bot}\not\!P_f}{M_f^2}]
+{\gamma}^{\mu}[a_5+a_6\frac{\not\!P_f}{M_f}+a_7\frac{\not\!q'_{\bot}}{M_f}+
a_8\frac{\not\!P_f\not\!q'_{\bot}}{M_f^2}]\}\,,
\end{eqnarray}
with
\begin{eqnarray}
&\displaystyle a_1=\frac{q'^2_{\bot}}{2M_fm'_1}n_1
+\frac{(f'_5(\vec{q'})w'_2-f'_6(\vec{q'})m'_2)M_f}{2m'_1w'_2}\,,\ \
\displaystyle a_2=\frac{(f'_6(\vec{q'})w'_2-f'_5(\vec{q'})m'_2)M_f}{2m'_1w'_2}\,,\nonumber\\
&\displaystyle
a_3=\frac{1}{2}n_1+\frac{f'_6(\vec{q'})M_f^2}{2m'_1w'_2}\,,\;\;\;\;
a_4=\frac{1}{2}(-\frac{w'_1}{m'_1})n_1+\frac{f'_5(\vec{q'})M_f^2}{2m'_1w'_2}\,,\nonumber\\
&\displaystyle
a_5=\frac{M_f}{2}n_2,a_6=\frac{M_fm'_1}{2w'_1}n_2\,,\;\;
a_7=0\,,\;\;
a_8=\frac{M_f^2}{2w'_1}n_2\,,\nonumber\\
&\displaystyle
n_1=\frac{1}{2}(f'_3(\vec{q'})+f'_4(\vec{q'})\frac{m'_1}{w'_1})\,,\;\;\;
n_2=\frac{1}{2}(f'_5(\vec{q'})-f'_6(\vec{q'})\frac{w'_1}{m'_1}).\nonumber
\end{eqnarray}

With Eq.~(\ref{previous}), the form factors may be presented by
overlapping integrations:
\begin{eqnarray}
&\displaystyle
k=\int\frac{d^3q}{(2\pi)^3}\frac{4b_1}{M_f^2(M+M_f)}\left[
\frac{q^2}{2}(\cos^2\theta-1)\right.\nonumber\\
&\displaystyle
\left(-{\alpha}_{11}E_f(M_f(-a_3-a_2b_3E_f+a_1b_4M_f)
+a_4(b_2E_f+b_4\vec{q}\cdot\vec{P_f}))\right.\nonumber\\
&\displaystyle
-({\alpha}_{11}a_8b_3E_f^2
-a_5b_3M_f^2+a_8b_3\vec{q}\cdot\vec{P_f})-{\alpha}_{11}E_f(M_f(-a_3-a_2b_3E_f+a_1b_4M_f)\nonumber\\
&\displaystyle
\left.+a_4(b_2E_f+b_4\vec{q}\cdot\vec{P_f})
+2a_8b_3E_f)\right)\nonumber\\
&\displaystyle -E_f({\alpha}_{11}-\frac{q\cos\theta}{|\vec{P_f}|})
\left(-a_5M_f^2+a_8b_3E_fq^2+a_6b_2E_fM_f
+a_6b_4M_f\vec{q}\cdot\vec{P_f}
-{\alpha}_{11}a_8b_3E_f\vec{q}\cdot\vec{P_f}\right)\nonumber\\
&\displaystyle
-\frac{E_fq^3\cos\theta}{|\vec{P_f}|}(1-\cos^2\theta)
\left((M_f(-a_3-a_2b_3E_f+a_1b_4M_f)\right.\nonumber\\
&\displaystyle\left.+a_4(b_2E_f+b_4\vec{q}\cdot\vec{P_f})+2a_8b_3E_f))\left.\right)
\right],
\end{eqnarray}
\begin{eqnarray}
&\displaystyle c_1=\int\frac{d^3q}{(2\pi)^3}
\frac{4b_1M}{M_f^2}\left[{\alpha}_{11}\frac{E_f}{M}
\left({\alpha}_{11}\frac{E_f}{M}(-{\alpha}_{11}a_4b_2E_f^2
-{\alpha}_{11}a_4b_4E_f\vec{q}\cdot\vec{P_f}\right.\right.\nonumber\\
&\displaystyle
+a_1b_2M_f^2+a_2b_3M_f\vec{q}\cdot\vec{P_f}
+a_3M_f(b_3q^2+{\alpha}_{11}E_f-{\alpha}_{11}b_3\vec{q}\cdot\vec{P_f}))
\left.-2{\alpha}_{11}a_8b_3E_f\vec{q}\cdot\vec{P_f}/M\right)\nonumber\\
&\displaystyle -\frac{E_fq\cos\theta}{M|\vec{P_f}|}
\left(-M({\alpha}_{11}\frac{E_f}{M})^2
(M_f(-a_3-a_2b_3E_f+a_1b_4M_f)+a_4(b_2E_f+b_4\vec{q}\cdot\vec{P_f}))\right.\nonumber\\
&\displaystyle-{\alpha}_{11}\frac{E_f}{M}({\alpha}_{11}a_8b_3E_f^2
-a_5b_3M_f^2+a_8b_3\vec{q}\cdot\vec{P_f})\nonumber\\
&\displaystyle+2{\alpha}_{11}\frac{E_f}{M}(-{\alpha}_{11}a_4b_2E_f^2-{\alpha}_{11}a_4b_4E_f\vec{q}\cdot\vec{P_f}
+a_1b_2M_f^2+a_2b_3M_f\vec{q}\cdot\vec{P_f}\nonumber\\
&\displaystyle
+a_3M_f(b_3q^2+{\alpha}_{11}E_f-{\alpha}_{11}b_3\vec{q}\cdot\vec{P_f}))
-{\alpha}_{11}2a_8b_3E_f\vec{q}\cdot\vec{P_f}/M\nonumber\\
&\displaystyle
\left.-{\alpha}_{11}\frac{E_f}{M}({\alpha}_{11}a_8b_3E_f^2
+a_5b_3M_f^2+a_8b_3\vec{q}\cdot\vec{P_f})\right)\nonumber\\
&\displaystyle
+\frac{q^2}{2M^2|\vec{P_f}|^2}(-M_f^2+(2E_f^2+M_f^2)\cos^2\theta)\nonumber\\
&\displaystyle
\left(M(a_4b_4q^2+a_2M_f+{\alpha}_{11}a_4b_2E_f-{\alpha}_{11}a_3M_f
-a_8-a_6b_4M_f+{\alpha}_{11}a_8b_3E_f)\right.\nonumber\\
&\displaystyle
-{\alpha}_{11}E_f(M_f(-a_3-a_2b_3E_f+a_1b_4M_f)+a_4(b_2E_f+b_4\vec{q}\cdot\vec{P_f}))\nonumber\\
&\displaystyle
-({\alpha}_{11}a_8b_3E_f^2
-a_5b_3M_f^2+a_8b_3\vec{q}\cdot\vec{P_f})\nonumber\\
&\displaystyle
-{\alpha}_{11}E_f
(M_f(-a_3-a_2b_3E_f+a_1b_4M_f)+a_4(b_2E_f+b_4\vec{q}\cdot\vec{P_f})\nonumber\\
&\displaystyle
\left.+2a_8b_3E_f))\right)\
-\frac{q^3E_f\cos\theta}{2M^2|\vec{P_f}|^3}(3M_f^2-(2E_f^2+3M_f^2)\cos^2\theta)\nonumber\\
&\displaystyle
\left.\left((M_f(-a_3-a_2b_3E_f+a_1b_4M_f)+a_4(b_2E_f+b_4\vec{q}\cdot\vec{P_f})+2a_8b_3E_f)\right)\right],
\end{eqnarray}
\begin{eqnarray}
&\displaystyle c_2=\int\frac{d^3q}{(2\pi)^3}\frac{4b_1M}{M_f^2}
\left[{\alpha}_{11}\frac{E_f}{M}\left({\alpha}_{11}E_f
(a_4b_4q^2+a_2M_f+{\alpha}_{11}a_4b_2E_f-{\alpha}_{11}a_3M_f)\right.\right.\nonumber  \\
&\displaystyle
\left.-(-a_8b_3q^2-a_6b_2M_f+{\alpha}_{11}a_8E_f)\right)\nonumber \\
&\displaystyle +\frac{E_fq\cos\theta}{M|\vec{P_f}|}
\left((-{\alpha}_{11}^2E_f)(M_f(-a_3-a_2b_3E_f+a_1b_4M_f)+a_4(b_2E_f+b_4\vec{q}\cdot\vec{P_f}))\right.\nonumber\\
&\displaystyle -{\alpha}_{11}({\alpha}_{11}a_8b_3E_f^2
-a_5b_3M_f^2a_8b_3\vec{q}\cdot\vec{P_f})-{\alpha}_{11}E_f
(a_4b_4q^2+a_2M_f+{\alpha}_{11}a_4b_2E_f-{\alpha}_{11}a_3M_f)\nonumber\\
&\displaystyle
+(-a_8b_3q^2-a_6b_2M_f+{\alpha}_{11}a_8E_f)\nonumber\\
&\displaystyle
-{\alpha}_{11}E_f(a_4b_4q^2+a_2M_f+{\alpha}_{11}a_4b_2E_f-{\alpha}_{11}a_3M_f
\left.-a_8-a_6b_4M_f+{\alpha}_{11}a_8b_3E_f)\right)\nonumber\\
&\displaystyle
+\frac{q^2}{2M|\vec{P_f}|^2}(-M_f^2+(2E_f^2+M_f^2)\cos^2\theta)\left
(a_4b_4q^2+a_2M_f+{\alpha}_{11}a_4b_2E_f\right.\nonumber\\
&\displaystyle
\left.-{\alpha}_{11}a_3M_f-a_8-a_6b_4M_f+{\alpha}_{11}a_8b_3E_f\right)
-\frac{q^2E_f}{2M|\vec{P_f}|^2}(3\cos^2\theta-1)\nonumber\\
&\displaystyle
\left(-{\alpha}_{11}E_f(M_f(-a_3-a_2b_3E_f+a_1b_4M_f)+a_4(b_2E_f+b_4\vec{q}\cdot\vec{P_f}))\right.\nonumber\\
&\displaystyle -({\alpha}_{11}a_8b_3E_f^2
-a_5b_3M_f^2+a_8b_3\vec{q}\cdot\vec{P_f})-{\alpha}_{11}E_f(M_f(-a_3-a_2b_3E_f+a_1b_4M_f)\nonumber\\
&\displaystyle
\left.+a_4(b_2E_f+b_4\vec{q}\cdot\vec{P_f})+2a_8b_3E_f))\right)
-\frac{q^3\cos\theta}{2M|\vec{P_f}|^3}[(4E_f^2+M_f^2)\cos^2\theta-(2E_f^2+M_f^2)]\nonumber\\
&\displaystyle \left(\right.
(M_f(-a_3-a_2b_3E_f+a_1b_4M_f)+a_4(b_2E_f+b_4\vec{q}\cdot\vec{P_f})+2a_8b_3E_f))\left.\right)\left.\right],
\end{eqnarray}
\begin{eqnarray}
&\displaystyle h=\int\frac{d^3q}{(2\pi)^3}\frac{4b_1(M+M_f)}{M_f^2}
\left[(\frac{q\cos\theta}{|\vec{P_f}|}-{\alpha}_{11})\right.
\frac{E_f}{M}(a_8b_3q^2+a_6b_2M_f+{\alpha}_{11}a_8E_f)\nonumber\\
&\displaystyle +{\alpha}_{11}\frac{E_f^2q\cos\theta}{M|\vec{P_f}|}
(a_8-a_6b_4M_f+{\alpha}_{11}a_8b_3E_f)-{\alpha}_{11}\frac{E_fq\cos\theta}{M|\vec{P_f}|}
b_3({\alpha}_{11}a_8E_f^2+a_5M_f^2-a_8\vec{q}\cdot\vec{P_f})\nonumber\\
&\displaystyle
-\frac{q^2}{2M\vec{P_f}^2}(-M_f^2+(2E_f^2+M_f^2)\cos^2\theta)
(a_8-a_6b_4M_f+{\alpha}_{11}a_8b_3E_f)\nonumber\\
&\displaystyle
+\frac{q^2E_f}{2M\vec{P_f}^2}(3\cos^2\theta-1)
(b_3({\alpha}_{11}a_8E_f^2+a_5M_f^2-a_8\vec{q}\cdot\vec{P_f}))\nonumber\\
&\displaystyle
\frac{q}{2}(\cos^2\theta-1)
\left.\left(-{\alpha}_{11}\frac{E_f}{M}(2(b_3M_f(a_3{\alpha}_{11}-a_2)+a_4(b_2
+{\alpha}_{11}b_4E_f))+4a_8b_3)\right)\right.\nonumber\\
&\displaystyle
\left.+\frac{E_fq^3\cos\theta}{2M|\vec{P_f}|^2}(1-\cos^2\theta)
(2(b_3M_f(a_3{\alpha}_{11}-a_2)+a_4(b_2+{\alpha}_{11}b_4E_f))+4a_8b_3)\right].
 \end{eqnarray}

\end{document}